\documentclass[twocolumn,showpacs,preprintnumbers,amsmath,amssymb]{revtex4}

\usepackage[dvips]{graphicx}
\usepackage{dcolumn}
\usepackage{bm}
\newcommand{\nc}{\newcommand}
\def\nn{\nonumber\\}
\def\bea{\begin{eqnarray}}
\def\eea{\end{eqnarray}}
    
\nc{\braket}[1]{\langle\,{#1}\rangle}

\def\wt{\widetilde}

\def\pa{\partial}

\def\Z{\mathbf Z}  \def\R{\mathbf R}  

\def\V{{\cal V}}

\begin{document}

\preprint{RIKEN-TH-58}

\title{Low energy states of a semiflexible polymer chain with
  attraction and the whip-toroid transitions}

\author{Y. Ishimoto}
\affiliation{Theoretical Physics Laboratory, RIKEN, Wako 351-0198, Japan}
\author{N. Kikuchi}
\affiliation{Institut f\"ur Physik, Johannes Gutenberg-Universit\"at Mainz, Staudinger Weg 7, D-55099 Mainz, Germany}

\date{\today}
             
\begin{abstract}

We establish a general model for the
whip-toroid transitions of a semiflexible homopolymer chain using the path
integral method and the $O(3)$ nonlinear sigma model on a line segment with
the local inextensibility constraint. We exactly solve the energy levels of
classical solutions, and show that some of its classical configurations
exhibit toroidal forms, and the system has phase transitions from a whip to
toroidal states with a conformation parameter $c=\frac{W}{2 l} \left(
\frac{L}{2\pi} \right)^2$. We also discuss the stability of the toroid
states and propose the low-energy effective Green function. Finally, with
the finite size effect on the toroid states, predicted toroidal properties
are successfully compared to experimental results of DNA condensation.

\end{abstract}


\maketitle

\newpage
\section{Introduction}

In nature, biological macromolecules are often found in collapsed 
states \cite{DE86,DG79,GK94}. Proteins take unique three-dimensional 
conformations in the lowest energy state (native state), which is of 
great importance in its functionality \cite{FP02}. 
DNA in living cells is often packaged tightly, for instance, inside phage capsids. 
Recent advances in experimental techniques mean it is now possible to
study the conformational properties of biopolymers at single molecular
level \cite{PSLC95,SABBC96,PSC97,GHFA02,STSGAB01}. 
As well as its biochemical, medical, and industrial importance, (bio-)polymers have drawn much attention 
\cite{GK94,FP02,PSLC95,SABBC96,PSC97,GHFA02,STSGAB01,GS76,B91,B96,YYK98,CVH03,HD01}. 

To increase our understanding of their physical properties, a flexible
homopolymer chain in a dilute solution, as the simplest model, has been heavily
investigated \cite{DG79,DE86,GK94,LGK78,KF84,DG85,KTD96a,ALO02,KRPY05}. 
When the temperature is lowered, or the solvent quality is changed from good to poor, 
the resulting effective attractive interactions between monomers 
can cause the polymer to undergo a coil-globule transition (collapse transition)
from an extended coil to a compact globule state \cite{DG79,DE86,GK94}. 
Both equilibrium \cite{LGK78,KF84,DG79,DE86,GK94} and dynamical \cite{DG85,KTD96a,ALO02,KRPY05} properties of the coil-globule
transition of the flexible chain are now well understood.

However, many biological macromolecules such as DNA, F-actin, and
collagen show large persistence lengths and are classified as
semiflexible chains \cite{DE86,K04,KF72}. 
For instance, double stranded DNA in aqueous solution, mostly with segment diameter $\sigma\,{\simeq}\,2\,nm$, has the
persistence length $l\,{\simeq}\,50\!\sim\!60\,nm$. Therefore natural
DNAs behave as semiflexible chains when their contour lengths are several orders longer than $l$ \cite{DE86,K04,KF72}. 
In such cases in a poor solvent condition, the balance
between the bending stiffness and surface free energies induces
toroidal conformation rather than spherical globule of a flexible chain 
\cite{GK81,SIGPB03,HDB95,UO96,SMW00,SGM02,PW00,MKPW05,PHG98,TYYKK05}. 
In fact, when we put condensing agents as multivalent cations into DNA
solution, it can cause DNA to undergo the condensation from a
worm-like chain (whip or coil) to toroidal states 
\cite{GS76,B91,B96,YYK98}.

Towards the understanding of the ``whip(or coil)-toroid transition'' of a
 semiflexible homopolymer chain, or of a DNA chain, many experimental
 and theoretical works have been done, in particular, in a poor solvent condition \cite{MPM04,CW04,NSKY96,NY98,KT99,KTD96,IPB98,MSIMPB05,GK81,SIGPB03,HDB95,UO96,SMW00,SGM02,PW00,MKPW05,PHG98,TYYKK05,EM78,PAB90,FH99,B98}.
Extensive results from experiments showed that collapsed DNA exists in toroid, rod, sphere and spool-like phases with the toroid being the
most probable \cite{NSKY96,EM78,PAB90,FH99,B98}.
Simulations using Monte Carlo, Langevin approaches or Gaussian variational method, 
calculated phase diagram for the semiflexible chain in a poor solvent 
\cite{MPM04,CW04,NSKY96,NY98,KT99,KTD96,IPB98,MSIMPB05}.
In theoretical works, existing phenomenological models balance the 
bending and surface free energies to estimate toroidal properties 
\cite{GK81,SIGPB03,HDB95,UO96,SMW00,SGM02,PW00,MKPW05,PHG98,TYYKK05}. 
It becomes increasingly probable that toroid is the stable lowest
energy state --- the ground state.

We note, however, that the theoretical aspects of the works assume a priori toroidal geometry as the stable lowest energy state with no theoretical proof 
\cite{SMW00}. 
Moreover, compared to the theory of coil-globule 
transition of a flexible chain \cite{LGK78,KF84,DG79,DE86,GK94}, 
which are well described by Gaussian approximation and field theoretical formalism 
\cite{LGK78,KF84}, 
there is no simple ``microscopic'' theory, which contains the salient physics to demonstrate the whip-toroid transition of the semiflexible polymer. 

Difficulties in formulating theory results specifically from the ``local
inextensibility constraint'' of the semiflexible chain, which
  makes the theory non-Gaussian 
\cite{K04}, 
and also from the ``non-local nature'' of the attractive interaction along
the polymer chain, which makes the theory analytically intractable. As a result,
even for the simplest semiflexible chain model without attraction, i.e. the Hamiltonian (\ref{FreebendingHamiltonian}), only a few equilibrium properties are
analytically tractable such as the mean square end-to-end distance 
$\braket{{\bf R}^2}$ of a free chain \cite{K04,HK05} and that of a semiflexible chain confined to a spherical surface \cite{SW03}.

To overcome these problems, we propose a microscopic model to describe the
whip-toroid transitions of a semiflexible homopolymer chain at low energy --- at low temperature or at large persistence length. To explore the equilibrium distribution (Green function) of a semiflexible chain, the path integral formulation is applied rather conventionally. Note that a semiflexible homopolymer chain in equilibrium at low energy satisfies
the local inextensibility constraint. Also, if the chain satisfies the local
inextensibility constraint, its Hamiltonian becomes equivalent to the O(3) nonlinear sigma model on a line segment. Therefore, a semiflexible homopolymer chain at low energy can be formulated in the path integral of the O(3) nonlinear sigma model on a
line segment. It is the first time that the local inextensibility constraint and the non-local attraction in the path integral are employed together and are solved clearly. Exploring it in detail, we find the toroid states as the ground state and the whip-toroid transitions of the semiflexible chain at low energy, which can also be found in our preprint \cite{IK051}. We then discuss and test the stability of the toroidal solutions, and propose the low-energy effective Green function. We show, in final sections, that our predictions on toroidal properties are in sufficiently quantitative agreement with the experiments \cite{YYK98,B96}.

The paper is organized as follows. In sections II and III, 
a semiflexible polymer chain with a delta-function attractive potential is formulated in the path integral method. We then deduce $O(3)$ nonlinear sigma model on a line segment with the local inextensibility constraint. 
In section IV, we derive the classical equations of motion for the
nonlinear sigma model action, and solve them explicitly. We also prove
that our solutions represent the general solutions of the
equations. The precise microscopic Hamiltonian, or the energy levels,
are obtained from the solutions, and the conditions for the stable
toroids are given. We also investigate the phase transitions in the
presence of the attractive interactions. 
Section V is devoted to the stability of the toroidal states under the `quantum' fluctuations away from classical solutions. We also construct the low-energy effective Green function from those of the whip and toroid states using perturbation theory.
In section VI, the finite size effect is introduced and the theory is mapped
onto physical systems. Assuming the hexagonally packed cross sections 
and van der Waals interactions, 
we show that our microscopic model does fit well quantitatively with a macroscopic property of the toroids --- the mean toroidal radius in the experiments \cite{YYK98,B96}. 
In the final section, our conclusion summarizes the paper and discussions are given with respect to the literature and the future prospects. Note the precise definition of the delta function potential is given in Appendix A, and the $SO(3)$ transformations are described in Appendix B.

\section{Polymer chain as a line segment}

In the continuum limit, the Green function (end-to-end distribution) of a semiflexible polymer chain
with attractive interactions can be given by the path integral: 
\bea
G(\vec{0},\vec{R};\vec{u}_i,\vec{u}_f;L,W)=
{\cal N}^{-1}\!\!\!\! \int_{\vec{r}(0)=\vec{0}, {\vec{u}(0)=\vec{u}_i}}^{\vec{r}(L)=\vec{R}, {\vec{u}(L)=\vec{u}_f}} 
  \!\!\!\!\!\!\!\!\!\!\!\!\!\!\!\!\!\!\!\!\!\!\!\!\! {\cal D} [\vec{r}(s)]\, e^{- {\cal H}[\vec{r}, \vec{u}, W]}
\eea
with the local inextensibility constraint $|\vec{u}|^2 = 1$ 
\cite{K04,KF72}. 
$s$ is the proper time along the semiflexible polymer chain of total contour length $L$. $\vec{r}(s)$ denotes the pointing vector at the `time' $s$ in our three dimensional space while
$\vec{u}(s)\equiv\frac{\pa \vec{r}(s)}{\pa s}$ corresponds to the unit bond (or tangent) vector at $s$. ${\cal N}$ is the normalisation
constant (\ref{normalizationN}).

Following Freed et al. and Kleinert \cite{KF84,K04}, the dimensionless Hamiltonian can be
written by 
\bea
  {\cal H}[\vec{r}, \vec{u}, W] 
  &=& \int_0^L ds\, \left[ H(s) + V_{AT}(s) \right]
\eea
where $H(s)$ and $V_{AT}(s)$ are the local free Hamiltonian and the
attractive interaction term, respectively:
\bea
H(s) &=& \frac{l}{2} \left\vert \frac{\pa}{\pa s} \vec{u}(s)
\right\vert^2\!\!\!,\label{FreebendingHamiltonian}\\
V_{AT}(s) &=& - W \int_0^s d s^\prime \delta \left( \vec{r}(s) - \vec{r}(s^\prime) \right)\!.
\label{VDWhamiltonian}
\eea
$l$ is the persistence length and $W$ is a positive coupling constant
of the attractive interaction between polymer segments. Thermodynamic
$\beta=1/(k_BT)$ is implicitly included in $l$ and $W$, which can be
revived when we consider the thermodynamic behaviours of the system. $l$ is assumed to be large enough to realise its stiffness: $l \gg l_b$, where $l_b$ is the bond length.
Note that there has been no consensus about the form of attractions, 
but people in the literature agree that effective attractions derive the toroidal geometry \cite{GNS02}. 
For example, in DNA condensations, interplay between charges, salt and
other unsettled (unknown) elements derives extraordinary
  short-range dominant effective attraction in a poor solvent
  condition. Therefore, we introduce the above delta-function
    potential $V_{AT}(s)$ for the modelling of the DNA
    condensation in a poor solvent condition, again as in Freed et
    al. \cite{KF84,K04}. As you can read off from the above,
    $V_{AT}(s)$ takes the non-local form, since the form at $s$
    contains information at the other points $s^\prime \in (0,s)$. In
    $V_{AT}(s)$, we omit the symbol for the absolute value
    $|\vec{r}(s)-\vec{r}(s^\prime)|$. (see Appendix A for the precise definition of the potential.)

In what follows, we express
$\vec{r}$ by the unit bond vector $\vec{u}$ and therefore 
the Hamiltonian ${\cal H}(\vec{u})$ in terms of 
$\vec{u}$. Hence, the Green function $G\left( \vec{0}, \vec{R};
\vec{u}_i, \vec{u}_f; L, W \right)$ becomes a path integral over $\vec{u}$ with the positive coupling constant $W$, regardless of $\vec{r}$,
\bea
G = \int_{\vec{u}_i}^{\vec{u}_f} {\cal D} [\vec{u}(s)]\, 
    \delta\left({\textstyle \int_0^L ds\, \vec{u}(s) - \vec{R} }\right) \, e^{- {\cal H}[\vec{u}, W]} ,
\eea
where we used $\vec{r}(L)= \int_0^L ds\, \vec{u}(s)$ 
and the Jacobian is absorbed by ${\cal N}$ which is neglected here. 
The delta function selects out the end-to-end vector. Basic properties of the Green function is given below.

Due to the local
inextensibility constraint $|\vec{u}(s)|^2=|\pa \vec{r}(s)|^2=1$, the total length of the polymer chain is strictly $L$ for
$G(\vec{R},L,W)$. Thus, the Green function as a distribution function exhibits a hard shell at $|\vec{R}| =L$:
\bea
 G\left( \vec{0}, \vec{R}; \vec{u}_i, \vec{u}_f; L, W \right) &=& 0 \quad {\rm for~} |\vec{R}|>L .
\eea
That is
\bea
 \int_{|\vec{R}|\leq L} d^3 \vec{R} \; G\left( 0, \vec{R}; \vec{u}_i, \vec{u}_f; L, W \right) = 1 .
\eea
It further means that the normalisation constant is given by
\bea
  {\cal N} = \int_{|\vec{R}|\leq L}\!\!\!\!\!\!\!\!\!\!\!\!d^3 \vec{R} \; \int_{\vec{u}_i}^{\vec{u}_f}\!\!\! {\cal D} [\vec{u}(s)]\, 
    \delta\left({\textstyle \int_0^L ds\, \vec{u}(s) - \vec{R} }\right) \, e^{- {\cal H}[\vec{u}, W]} .\label{normalizationN}
\eea

\section{O(3) nonlinear sigma model on a line segment}

When $W=0$, our free dimensionless Hamiltonian is given solely by $\vec{u}$ field:
\bea
  {\cal H}(\vec{u}) &\equiv& {\cal H}(\vec{r},\vec{u},W=0)
  \nn
  &=& \frac{l}{2} \int_0^L ds\, \left| \pa \vec{u}(s) \right|^2
\label{dimlessuhamil}
\eea
with the constraint $|\vec{u}(s)|^2=1$. This can be interpreted as the low energy limit of a
linear sigma model on a line segment, or quantum equivalently 
a nonlinear sigma model on a 
line segment, rather than some constrained Hamiltonian system. 

In this section we consider $O(3)$ nonlinear sigma model on a line
segment for the path integral formulation of the semiflexible polymer chain.
This is nothing but a quantum mechanics of a limited time 
$s\in [0,L]$ with a constraint. 
The constraint $|\vec{u}|^2 = 1$ restricts the value of $\vec{u}$ on a 
unit sphere $S^2$. This can be transformed into $u_3^2 = 1- u_1^2 -u_2^2.$

Substituting this into eq.(\ref{dimlessuhamil}) gives
\bea
  S[u_1,u_2] &=& \frac{l}{2} \int_0^L ds\; \left[ G^{ij} \right] \pa u_i(s) \pa u_j(s)
\eea
where the metric $G^{ij}$ on the unit sphere in three dimensional $\vec{u}$-space
\bea
G^{ij} [u_1, u_2] &\equiv& \left( 
   \begin{array}{cc} 
   \frac{1-u_2^2}{1-\left( u_1^2+u_2^2 \right)} & \frac{u_1u_2}{1-\left( u_1^2+u_2^2 \right)}\\
   \frac{u_1u_2}{1-\left( u_1^2+u_2^2 \right)} & \frac{1-u_1^2}{1-\left( u_1^2+u_2^2 \right)} 
   \end{array} \right).
\eea
This is called the nonlinear sigma model since the action is $O(3)$
symmetric but some of its transformations are realised nonlinearly on
this $\{u_i\}$ basis. It is also equivalent to the classical Heisenberg model with a constraint of unit length spins $\vec{S_{I}}^{\!2}=1$ in the continuum limit \cite{CL95}.

The action can also be expressed in the polar coordinate:
\begin{eqnarray}
\left\{
\begin{array}{rcl}
  u_1 &=& r_u \sin \theta_u \cos \varphi_u 
  \\
  u_2 &=& r_u \sin \theta_u \sin \varphi_u 
  \\
  u_3 &=& r_u \cos \theta_u
\end{array}
\right.
\Leftrightarrow
\left\{
\begin{array}{rcl}
  r_u &=& |\vec{u}| 
  \\
  \theta_u &=& \arccos{\frac{u_3}{r_u}}
  \\
  \varphi_u &=& \arctan{\frac{u_2}{u_1}}
\label{polarcoordinatesu}
\end{array}
\right. ,
\end{eqnarray}
\bea
 S[\theta_u, \varphi_u] &=& \frac{l}{2} \int_0^L ds\; \left[ (\pa \theta_u)^2 + \sin^2 \theta_u (\pa \varphi_u)^2 \right] 
  \nn
  &=& \frac{l}{2} \int_0^L ds\; [ \wt G^{ii}]\, \pa \theta_i(s) \pa\theta_i(s)
\label{nonlinearsigmaaction}
\eea
where $(\theta_1,\theta_2) \equiv (\theta_u,\varphi_u)$, and the metric $\wt G^{ij}$ is given by the diagonal matrix:
\bea
\wt G^{ij} [\theta_1,\theta_2] &\equiv& \left( 
   \begin{array}{cc} 
   1 & 0\\
   0 & \sin^2 \theta_1
   \end{array} \right).
\eea
This is essentially the same as $G^{ij}[u_i]$ since both are metrics
on the same sphere $S^2$. The $SO(3)$ transformations of the polar
coordinates $(\theta_u, \varphi_u)$ can be expressed by three
infinitesimal parameters $g_i$ (see Appendix B):
\bea
  \left\{ \begin{array}{rcl}
  \delta \theta_u &=& g_1 \sin \phi_u - g_2 \cos \phi_u ,
  \\
  \delta \phi_u 
  &=& \cot \theta_u \left( g_1 \cos \phi_u + g_2 \sin \phi_u  \right) - g_3 .
  \end{array} 
  \right. 
\eea

The canonical quantisation of the action
 (\ref{nonlinearsigmaaction}) is suitable for the investigation of the
 local nature of the system, but not for its global nature such as toroidal conformations. Therefore,
 we focus on the classical solutions of the action
 (\ref{nonlinearsigmaaction}) and consider the quantum fluctuations
 around the classical solutions using the path integral method.
Integrating the action (\ref{nonlinearsigmaaction}) by parts gives
\bea
 S[\theta_u, \varphi_u]\!&=&\!\!-\frac{l}{2} \int_0^L \!\!ds\!\!\; \left[
   \theta_u \pa^2 \theta_u + \varphi_u \left( \pa \circ \sin^2 \theta_u\circ\pa \right) \varphi_u \right]\nonumber\\
 & &+\Bigl[{\rm Surface} \Bigr]_0^L
 \label{parts}
\eea
where $\circ$ stands for the composition of the mappings and the surface term
$
 \Bigl[{\rm Surface} \Bigr]_0^L
 = \frac{l}{2} \Bigl[ \theta_u \pa \theta_u + \sin^2 \theta_u  \varphi_u \pa \varphi_u \Bigr]_0^L .
$
The surface term might be neglected by taking the north pole of the
polar coordinates $(\theta_u(0), \varphi_u(0))=(0,0)$ and considering
the static solutions. By setting the north pole, half of the surface
term vanishes. Given that we have the static solutions,
i.e. $\vec{u}(s)\sim \braket{\vec{u}\,}$ 
the surface term
contribution becomes much smaller compared to the bulk term
${\sim}O\left(\frac{l_b}{L}\right)$ where $l_b$ is the constant bond
length. Minimizing the action (\ref{nonlinearsigmaaction}) in terms of $\theta_u$
and $\varphi_u$ yields the classical equations of motion:
\bea
 \left[ - \pa^2  + \frac{\sin 2\theta_u}{2 \theta_u} (\pa \varphi_u)^2 \right] \theta_u &=& 0
 \nn
 \left[ \,\pa^2 + 2 (\pa \theta_u) \cot \theta_u \pa\, \right] \varphi_u &=& 0 .
\label{EOM}
\eea

\section{Classical solutions and the whip-toroid transition}

Our aim in this section is to explore classical solutions of
eq.(\ref{EOM}) and to study the lowest energy states and the whip-toroid phase transition in the presence of attractive interactions.

\subsection{Classical solutions}

\begin{figure}[h]
\centering
\includegraphics[width=8cm]{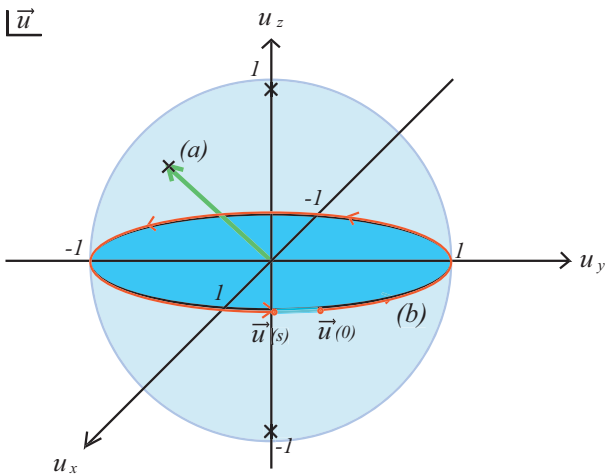}
 \caption{
Classical solutions of eq.(\ref{EOM}): (a) constant $\vec{u}$, (b) a path along a great circle on $S^2$.
}
      \label{fig:classical}
\end{figure}

Consider classical solutions of eq.(\ref{EOM}) with a trial solution
$\dot{\theta_u}=0$. The first equation of (\ref{EOM}) leads to $\sin 2\theta_u (\dot{\varphi_u})^2 =0 $.
Thus, the solution is either $\theta_u = 0, \frac{\pi}{2}, \pi$ or
$\dot{\varphi_u} = 0$. The solutions $\theta_u=0,\pi$ or
$\dot{\varphi_u}=0$ with $\dot{\theta_u}=0$ are equivalent to having a
constant $\vec{u}$. Accordingly, classical solutions reduce to $\theta_u =
\frac{\pi}{2}$ or $\vec{u}=const$. When we substitute $\theta_u =
\frac{\pi}{2}$ into the second equation of motion (\ref{EOM}), we obtain $\pa^2 \varphi_u =0$.
Therefore, we have the two classical solutions
\bea
  && \vec{u}(s) = const.
  \nn
  &&or~~
  \nn
  && \theta= \frac{\pi}{2} \quad and \quad \varphi_u = a s + b ,
  \label{classical sol}
\eea
where $a, b$ are constants. Note that the second
classical solution of eq.(\ref{classical sol}) is the uniform motion
of a free particle on the sphere (see Fig.\ref{fig:classical}).

By symmetry argument, we state that the solutions
(\ref{classical sol}) represent all the classical solutions. 
That is either a constant $\vec{u}(s)$ (rod solution)
or a rotation at a constant speed along a great circle 
on the $S^2$ (toroid solution).

\noindent
{\bf Proof)}

The theory has $O(3) \supset SO(3)$ global symmetry. Accordingly, one can take any initial value of $\vec{u}(0)$ for a classical solution. In other words, one may set $\vec{u}(0)$ to be the north pole for the representatives of the classical solutions, using two degrees of freedom of $SO(3)$ rotations.
In addition, by $SO(2)$ local rotation symmetry or by one residual degree of freedom of $SO(3)$, we can freely set the orientation of $\partial \vec{u}(0)$. For example, 
$\left( \dot{\theta_u}(0), \dot{\varphi_u}(0) \right)=(0, a)$ or $(a,0)$.
So, one may set the initial values as 
\begin{eqnarray}
  \vec{u}(0) &=& ( \theta_u(0), \varphi_u(0) ) = \left( \frac{\pi}{2}, 0 \right) ,
  \nn
  \partial \vec{u}(0) &=& \left( \dot{\theta_u}(0), \dot{\varphi_u}(0) \right)=(0, a) ,
\end{eqnarray}
with a condition $a\geq 0$. The non-negative real constant $a$ turns out to be the only degree of freedom that represents all the classical solutions.

Substituting these initial values to the equations of motion (\ref{EOM}), we obtain at $s=0$
\begin{eqnarray}
  \partial^2 \theta_u &=&0,\quad  \partial^2 \varphi_u =0 .
\label{keep}
\end{eqnarray}
So, an infinitesimal change $\epsilon$ of the variable $s$ yields
\begin{eqnarray}
  ( \theta_u(\epsilon), \varphi_u(\epsilon) ) = \left( \frac{\pi}{2}, \epsilon a \right), \quad 
  \!\!\left( \dot{\theta_u}(\epsilon), \dot{\varphi_u}(\epsilon) \right) = (0, a).
\end{eqnarray}
As one can see in the equations of motion and in the above, so long as $\theta_u(s)=\frac{\pi}{2}$, eq.(\ref{keep}) holds at any $s$.
Hence, the initial conditions leads to the pair of conditions, $\partial\theta_u(s)=0$ and $\theta_u(0)=\frac{\pi}{2}$. In other words, the pair of the conditions exhaust the representatives of the classical solutions. Thus, found solutions may well be regarded as the general solutions.
$(Q.E.D.)$

\medskip

Note that the solutions (\ref{classical sol}) can be regarded as `topological' solutions in a sense that they are solitonic solutions.

\subsection{Non-local attractive interactions as a topological term}

Now we consider the attractive interaction term (\ref{VDWhamiltonian}). 
It is difficult to interpret it in the context of quantum theory
due to its non-local nature along the polymer chain. 
However, we can solve them with our classical solutions (\ref{classical sol}).
Let us rewrite
eq.(\ref{VDWhamiltonian}) with 
\bea
\vec{r}(s) - \vec{r}(s^\prime) = \int_0^s dt\, \vec{u}(t) - \int_0^{s^\prime} dt\, \vec{u}(t) = \int_{s^\prime}^s dt\, \vec{u}(t),
\nn
\eea
that is,
\bea
V_{AT}(s) = -W \int_0^s d s^\prime\; \delta\left( \int_{s^\prime}^s dt\, \vec{u}(t) \right).
\eea
Hence the problem is now reduced to the one in the $\vec{u}$ space: finding non-zero values of
$\delta\left( \int_{s^\prime}^s dt\, \vec{u}(t) \right)$ with the
classical solutions (\ref{classical sol}). 
That is to find $\vec{u}(s^\prime)$ for a given $s$, which satisfies
$\left| \int_{s^\prime}^s dt\, \vec{u} \right| = 0$. Note that, exactly speaking, the integration over $s^\prime$ is from $0$ to $s-\epsilon$ with an infinitesimal positive constant $\epsilon$ (see Appendix A). Thus, we exclude the $s^\prime = s$ case in the following.

In the polar coordinates (\ref{polarcoordinatesu}), this is expressed by
\bea
   \int_{s^\prime}^s dt\, \sin \theta_u \cos \varphi_u &=& 0,
   \nn
   \int_{s^\prime}^s dt\, \sin \theta_u \sin \varphi_u &=& 0,
   \nn
   \int_{s^\prime}^s dt\, \cos \theta_u &=& 0.
\label{u-conditionforAttraction}
\eea
The first classical solution ($\vec{u}=const.$) does not satisfy these equations and thus derives no attractive interactions. If we
substitute the second classical solution of eq.(\ref{classical sol})
into eqs.(\ref{u-conditionforAttraction}), we have $\cos \theta_u(s)=0$,
\bea
   \int_{s^\prime}^s\!\!\!dt\, \cos (at\!+\!b) \!\!\!&=&\!\!\! \frac1a \left( \sin (as\!+\!b)\!-\!\sin (as^\prime\!\!+\!b) \right)\!=\!0,
   \nn
   \int_{s^\prime}^s\!\!\!dt\, \sin (at\!+\!b) \!\!\!&=&\!\!\! \frac1a \left( \cos (as^\prime\!\!+\!b)\!-\!\cos (as\!+\!b)\right)\!=\!0.
\eea
Hence we have solutions: $s-s^\prime = 2n\pi/a >0$, $n\in\Z$. 
Without any loss of generality, 
we assume $a>0$ and $n\in\Z_+$. Introducing $N(s) \equiv [as/2\pi]$ by Gauss' symbol
\footnote{
Gauss' symbol $[x]$ gives the greatest integer that is not exceeding $x$.
}, we obtain
\bea
& &\int_{s-2\pi/a}^s dt\, \vec{u}(t) = \int_{s-4\pi/a}^s dt\,
  \vec{u}(t) \nonumber\\
& &= \cdots = \int_{s- 2\pi N(s)/a}^s dt\, \vec{u}(t) = 0.
\eea
Therefore, the attractive potential is given by
\bea
  V_{AT}(s) = - W \cdot N(s). 
\eea
Note that $N(L)$ represents the winding number of the classical solution
(\ref{classical sol}) along a great circle of $S^2$ (see
Fig.\ref{fig:classical}). Finally, an integration over $s$ yields the dimensionless Hamiltonian 
with our classical solutions:
\bea
  &{\cal H}&[\vec{u},W]
  = \int_0^L ds\, H(s) + \int_0^L ds\,V_{AT}(s)
  \nn
  &=& \frac{Ll}{2} a^2 - W \left[ \frac{2\pi}{a} \sum_{k=1}^{N(L)-1} k + \frac{2\pi}{a} \left( \frac{aL}{2\pi} - N(L) \right) N(L) \right]
  \nn
   &=& \frac{Ll}{2} a^2
   - W L\cdot N(L) \left\{ 1-\frac{\pi}{aL}\left(N(L)+1\right) \right\}
   \label{Hamilwithattraction}
   .
\eea
The first term denotes the bending energy, 
and the second and the third terms are thought of as 
`topological' terms from the winding number. When the chain of contour length $L$ winds $N(L)$ times we have the
$N(L)$ circles of each length $\frac{2\pi}{a}$ and the rest $\left(
L-\frac{2\pi}{a}N(L) \right)$. The second and third terms in the second line of
eq.(\ref{Hamilwithattraction}) result from the former and the latter
respectively. 

\subsection{The toroid and whip states}
\label{toroid and whip states}

\begin{figure}[h]
\centering
\includegraphics[width=8cm]{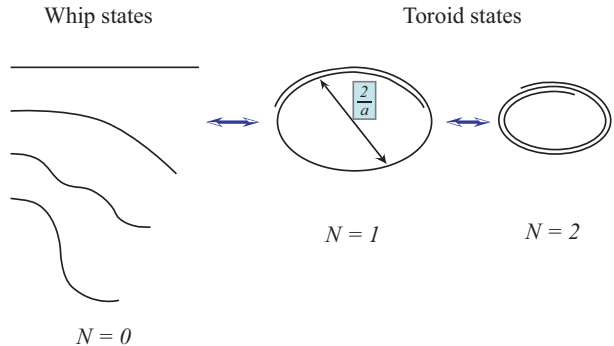}
 \caption{
 The whip ($N=0$) and toroid states ($N\geq 1$). The value of $b$ is given by the initial value of the bond vector $\vec{u}_i$.
}
      \label{fig:toroid-whip}
\end{figure}

The non-zero winding number of the classical solution in the $\vec{u}$
space means that the polymer chain winds in the $\vec{r}$ space as
well. That is, when $a>\frac{2\pi}{L}$, configurations around the
second classical solution (\ref{classical sol}) start forming a toroidal shape since
\bea
 \vec{r}(s) = \left( \begin{array}{c} 
    \frac{1}{a} \left\{ \sin (as +b)-\sin (b) \right\} \\
    -\frac{1}{a} \left\{ \cos (as +b)-\cos (b) \right\} \\
    const.
  \end{array} \right) ,
\eea
and stabilise itself by attracting neighbouring segments. We call such classical solutions the ``toroid states.''
Whenever $a$ increases and passes through the point
$\frac{2\pi n}{L}$ for $n\in\Z_+$, another toroid state
appears with the increased winding number $n$. Note that the radius of
the toroid state is given by $\frac{1}{a}$ (see Fig.\
\ref{fig:toroid-whip}). 
When $0<a\leq \frac{2\pi}{L}$, the chain cannot
wind like the toroid states. Both ends of the chain are not
connected to each other, thus can move freely as well as any other
parts of the chain fluctuate. 
As long as the total energy of the chain does not exceed the bending energy of $\frac{2\pi^2 l}{L}$ at {}$a=\frac{2\pi}{L}$, 
they can whip with zero winding number.

We call such low-energy extended coil states the ``whip states.'' 
Although the definition includes fluctuations around the classical solutions, unless otherwise stated, 
we primarily 
refer to the classical solutions of such states, which 
are rather bowstrings than whips.

In the next subsection we explore the exact energy levels of the whip and
toroid states, and discuss the phase transitions between these states.

\subsection{Favoured vacuum and toroid-whip transition}

The dimensionless Hamiltonian of the second classical solution
(\ref{classical sol}) is a function of $l, L, W$ and $a$: 
\bea
{\cal H}_{cl}(\!a,\!l,\!L,\!W\!)\!\equiv\!\frac{Ll}{2} a^2\!+\!\frac{\pi W }{a} N\!(\!L\!)\!\left(N\!(\!L\!)\!+\!1\right)\!-\!W L\!\cdot\!N\!(\!L\!).
 \label{Hamiltonian}
 \label{Hamiltoniancl}
\eea
This matches with the first classical solution when $\frac{N(L)}{a}=0$ for $a=0$ is defined. 
Accordingly, 
the above expression is valid for all classical solutions. Note 
that, since previous works assume a priori toroidal shape, no one 
clearly derived the precise microscopic Hamiltonian. Thus, we are now 
in a position to investigate exact energy levels of the whip and toroid states.

\begin{figure}[h]
\centering
\includegraphics[width=8cm]{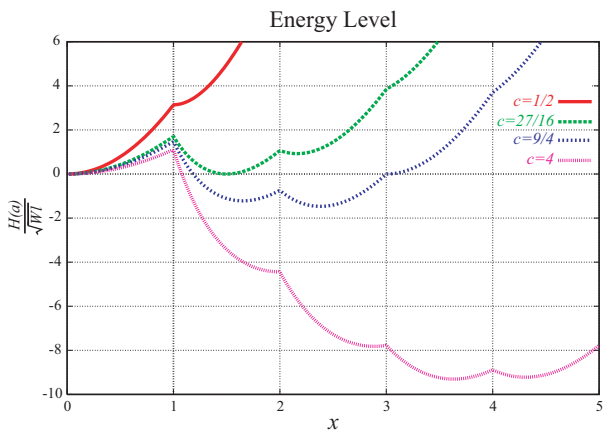} 
 \caption{\baselineskip=20pt
The dependence of the energy ${\cal H}(a)$ on $x=a L/2\pi$ and $c$. ${\cal H}(a)$ is scaled by the factor of $\sqrt{Wl}$ for convenience.
}
      \label{fig:energy level}
\end{figure}

Consider first a case with $L$, $W$, and $l$ fixed.
By definition, 
${\cal H}(a) \equiv {\cal H}_{cl}(a,l,L,W)$ is continuous in the
entire region of $a\geq 0$ and is a smooth function in each segment: 
\bea
a\in \left[ \frac{2\pi N}{L}, \frac{2\pi(N+1)}{L} \right] \quad {\rm for~} N\in
\Z_{\geq 0}.\label{inequalityofsegments}
\eea
However, it is not
smooth at each joint of the segments: $\frac{aL}{2\pi} \in \Z_+$. 
Introducing a new parameter 
$c\equiv \left( \frac{L}{2\pi}\right)^{\!2}\! \frac{W}{2l}$ out of three existing degrees of freedom,
we plot in Fig.\ref{fig:energy level} the energy levels as a function of
$a$ for different values of $c$, showing qualitative agreement with
Conwell et al. for the condensation of {\it $3$kb DNA} in various
  salt solutions \cite{CVH03}.
Note that, in what follows, 
we call the segment (\ref{inequalityofsegments}) the ``N-th segment''
counting from $0$-th, and we also call $c$ the ``conformation
parameter'' because the parameter $c$ solely determines the shape of this curve.

Suppose $N(L)=N$ is fixed, the Hamiltonian (\ref{Hamiltoniancl}) takes a
minimum at $a=a_c(N)\equiv\left( \frac{\pi W}{Ll}N(N+1)
\right)^{1/3}$. Accordingly, each segment falls into one of the following three cases: 
\begin{itemize}
\item[(i)] When $a_c(N) \leq \frac{2\pi N}{L}$, ${\cal H}(a)$ is a monotonic
function in the segment and takes its minimum at $a= \frac{2\pi N}{L}$. 
\item[(ii)] When $\frac{2\pi N}{L} <a_c(N)< \frac{2\pi (N+1)}{L}$, ${\cal H}(a)$ behaves quadratic in $a$ and takes its minimum at $a=a_c(N)$. 
\item[(iii)] When $\frac{2\pi (N+1)}{L}<a_c(N)$, ${\cal H}(a)$ is monotonic in the segment and takes its minimum at $a=\frac{2\pi (N+1)}{L}$.
\end{itemize}
The first and third cases are physically less relevant since they mean no (meta-)stable point in the segment. So, we focus on the second case.

\begin{figure}[h]
\centering
\includegraphics[width=8cm]{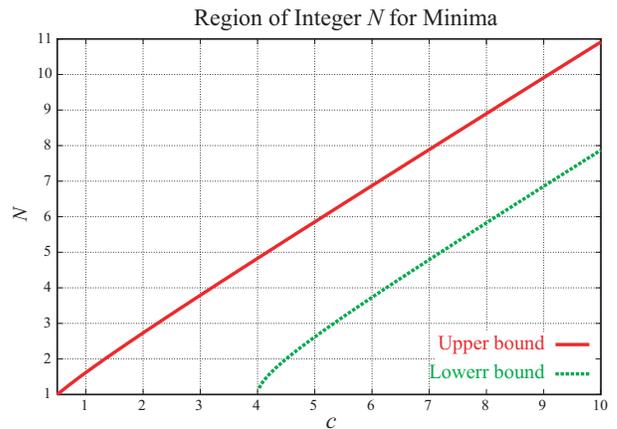}
 \caption{
The solid line is the upper bound and the dashed line is the lower bound of $N$ for the minima, {\it i.e.}, $N_{U,L}(c)$ (and $c_{U,L}^{(N)}$). The asymptotic values of $N_{U,L}(c)$ are both $N_{U,L}(c) \sim c$ ($c_{U,L}^{(N)} \sim N$). 
}
      \label{fig:phase transition}
\end{figure}

The condition on $N$ for the second case turns out to be (see Fig.\ \ref{fig:phase transition})
\bea
N_L(c) < &N& < N_U(c)  \quad {\rm for~} c\geq 4,
\nn 
1\leq &N& < N_U(c)  \quad {\rm for~} 0\leq c < 4,
\label{Nbounds}
\eea
where 
\bea
  N_L(c) &\equiv& \frac{c}{2} \left( 1 - \frac{2}{c} + \sqrt{1-\frac{4}{c}} \right), \nonumber\\
  N_U(c) &\equiv& \frac{c}{2} \left( 1 + \sqrt{1+\frac{4}{c}} \right) .
\eea
Note that, by replacing $N$ with $N(L)= [aL/2\pi]$, one can read the condition on $a$ as well.

As one can see in Figs.\ \ref{fig:energy level}, \ref{fig:phase
  transition} , there are apparently more than one (meta-)stable
toroid states at most values of $c$. This is because the first term of
bending in eq.(\ref{Hamiltoniancl}) is monotonically increasing, while
the other two terms in eq.(\ref{Hamiltoniancl}) are decreasing 
but not smoothly. This non-smoothness and the balance between two
factors lead to multiple local minima and potential barriers between them.
The number of minima is roughly given by the width of the region for $N$, i.e., $N_U(c) - N_L(c)$. 
For example, when $c \geq 4$,
\bea
  N_U(c) - N_L(c) &=& 1+ \frac{c}{2} \left( \sqrt{1+\frac{4}{c}} - \sqrt{1-\frac{4}{c}} \right) 
  \nn
  &=& 1 + \frac{c}{2} \left( \sum_{k=0}^\infty \frac{(-1)^{k}
      \left(-\frac12\right)_{k}}{k!} \left(\frac{4}{c}\right)^{k}\right.
    \nonumber\\
& &\,\,\,\,\,\,\,\,\,\,\,\,\,\,\,\left.- \sum_{k=0}^\infty \frac{\left(-\frac12\right)_{k}}{k!} \left(\frac{4}{c}\right)^{k} \right)
  \nn
  &=& 3 + 2 \left( \sum_{k=1}^\infty \frac{\left(\frac12\right)_{2k}}{(2)_{2k}} \left(\frac{4}{c}\right)^{2k} \right)
  \nn
  &>& 3 ,
\label{Nregion}
\eea
where $(a)_k=a(a+1){\cdots}(a+k-1)$ is the Pochhammer
  symbol. Therefore, there are at least three minima with positive
  winding numbers greater than 1. When $0 < c < 4$, the condition of having more than three minima is $c>\frac94$. 
To summarise, when $c>\frac94$ there exist at least three minima with 
positive winding numbers. It might be helpful to mention that, if we 
introduce the finite size effect in section VI, the number of minima 
could be reduced in some cases. 

One can plot the critical value of $c$ where the minimum of the $N$-th segment emerges and vanishes.
The lower bound of the $N$-th segment is
\bea
  c_L^{(N)} = \frac{N^2}{N+1} < N ,
\eea
while the upper bound is
\bea
  c_U^{(N)} = \frac{(N+1)^2}{N} > N+1 .
\eea
So, when $c$ satisfies the following inequality relation:
\bea
  c_L^{(N)}\!\!< c < c_U^{(N)},
\label{c_inequality}
\eea 
the $N$-th segment has a minimal
and (meta-)stable point. For example, when $\frac12 < c < 4$, 
the first segment $a\in [\frac{2\pi}{L}, \frac{4\pi}{L}]$ (i.e. $N=1$) 
has a minimal point at $a=a_c(1)$.

Now we discuss the critical points of the conformation parameter $c$ at which the conformational transitions between states may occur.
When $N_U(c) \leq 1$ (i.e. $c \leq \frac12$), the second condition in eq.(\ref{Nbounds}) vanishes
and thus the whip states only survive at low energy. In this parameter
region, the $a=0$ rod state will be favoured as the ground state with
vanishing energy. Including `quantum' fluctuations around $a=0$, we
call this phase the whip phase. Successively, at the critical value of $c=\frac12$, the whip phase to
whip-toroid co-existence phase transition would occur. On the other
hand, when $c > \frac12$, there always exists at least one (meta-)
stable toroid state with positive winding number $N(L)$. As $c$ grows over $\frac12$, the local minimum in the first
  segment decreases from some positive value.
Finally, when the energy of the $N=1$ stable toroid state balances with the
  ground state of the whip state (i.e. ${\cal H}_{cl}=0$), 
the whip-dominant to toroid-dominant phase transition may occur. 
Such a value of $c$ is $27/16$. 
Since there is a potential barrier between the $a=0$ rod and the
$N=1$ stable toroid states, the transition is first order.
When $c>27/16$, the toroid states will dominate the action. The energy 
plot (Fig.\ \ref{fig:energy level}) clearly shows that the transitions 
between the toroid states are also first order, if any, as there 
exist potential barriers between two successive minima. Further discussions on the phase transitions will be given in the final section.

For later convenience, we rewrite the Hamiltonian (\ref{Hamiltonian}) and $a_c(N)$ in terms of $c$ and the new variable $x\equiv \frac{aL}{2\pi}$
\bea
& &{\cal H}_{cl}(a,l,L,W) = \frac{WL}{2} {\cal H}(c,x)\nonumber\\
& &= \sqrt{2 \pi^2 W l} \left\{ \sqrt{c}\; {\cal H}(c,x) \right\} 
  = \frac{4\pi^2 l}{L} c \; {\cal H}(c,x),
\eea
where
\bea
 {\cal H}(c,x) &=& \frac{x^2}{2 c} + \frac1x [x] ([x]+1) - 2[x].
\eea
Therefore, $[x]=N(L)$ and $x_c([x])=\frac{a_c(N) L}{2\pi} = \left\{ c\cdot [x]([x]+1)\right\}^{1/3}$.

\section{Stability, quantum fluctuations, and perturbations}

So far we have dealt with the classical solutions, which are derived
from the first derivative of the action. Thus, they may correspond
to the global/local minima of the action in the configuration space. However, 
the solutions are not necessarily stable 
unless we take into account the attraction, 
since the second derivative test of the action with $W=0$ gives the non-positive Hessian. That is to say, they seem to be saddle points.
\bea
  \left. \frac{\delta^2 S[\theta_u,\varphi_u]}{\delta \varphi_u^2} \right|_{W=0}
  &=& 0,
  \nn
  \left. \frac{\delta^2 S[\theta_u,\varphi_u]}{\delta \theta_u^2} \right|_{W=0}
  &=& \left( l \dot{\varphi}_u^2 \right) \cos 2\theta_u ,
  \nn
  \left. \frac{\delta^2 S[\theta_u,\varphi_u]}{\delta \theta_u \delta \varphi_u} \right|_{W=0}
  &=& 
  -l \left( 2 \dot{\theta}_u \dot{\varphi}_u \cos(2 \theta_u) + \ddot{\varphi}_u \sin(2 \theta_u) \right) ,
\nn
  \det H |_{W=0} &=& \left| \begin{array}{cc}
              \pa_\varphi^2 S & \pa_\varphi \pa_\theta S \\
              \pa_\theta \pa_\varphi S & \pa_\theta^2 S
              \end{array}
               \right|_{W=0}
  \leq 0
\eea
In fact, the general whip states do not need to live in a flat plane
in $\R^3$ whereas the classical whip state does. So, the
transitions between the classical and the non-classical whip states have
the flat directions, {\it i.e.}, they can be seamless without any change of energy. Therefore, the stability problem is to be treated carefully with and without attraction.

\subsection{Stability and quantum fluctuations with attraction}

When the attraction is turned on, the toroid states with the winding number of more than two may become extremely (meta-)stable under the quantum fluctuations away from the classical solutions. It is not easy to show that all such second derivatives of the action give positive values and therefore stabilise the states, 
since the interaction term contains a special function of the quantum variable $\vec{u}$. However, there is a much easier way to see the stability. 

Consider any small fluctuation of a segment $\delta l_s$ from such a state. It gives rise to an increase of the energy:
\bea
  \delta {\cal H}(a) \geq W \cdot \delta l_s.
\eea
\begin{figure}[h]
\centering
\includegraphics[width=8cm]{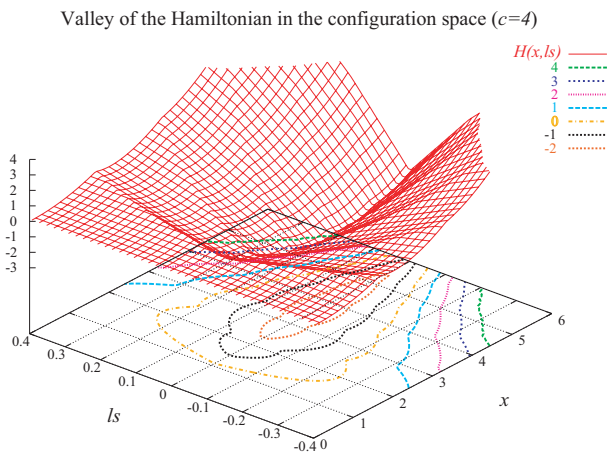}
 \caption{\small 
A sketch of the energy level ${\wt{\cal H}}(x,l_s)$ with the attraction. For ease of use, the maximum value is suppressed at $4$ and $l_s$ is normalised by $L$. The direction along the valley is parameterised by $a$ variable of the classical solution. Its perpendicular direction is the quantum fluctuation $l_s$ by $SO(2)$ away from the classical solutions. As one can see, $l_s$ is a flat direction for the whip states $(x<1)$.
}
      \label{fig:valley}
\end{figure}
More generally, we can write down the dimensionless Hamiltonian as a function of $a$, or $x$, and $l_s$, that is, the energy levels away from the toroid states.
$l_s$ is now defined as the length $l_s$ of the polymer segment shifted from an end of the toroid. The shifted polymer segment is locally rotated by $SO(2)$ transformation, for example, by $90$ degrees at each node, keeping the local bending energy unchanged.
\bea
  {\cal H}(a,l_s) &=& {\wt{\cal H}}(x,l_s)
  \nn
  &=& \frac{WL}{4 c} x^2 
  + \frac{WL}{2 x}  [x(1-\frac{l_s}{L})] 
    \left([x(1-\frac{l_s}{L})]+1 \right) \nn
  & &- W L [x(1-\frac{l_s}{L})] 
  + W l_s [x(1-\frac{l_s}{L})] ,
\eea
where $x=\frac{aL}{2\pi}$ (Fig.\ref{fig:valley}). 
The energy loss for the infinitesimal segment $\delta l_s$ is found to be
\bea
  \delta E = W N \delta l_s , 
\eea
where $N$ is the local number of the overlapped segments $N=[\frac{a(L-l_s)}{2\pi}]$. When $l_s=L$, the energy becomes $E= \frac{Ll}{2} a^2$ of the bending energy.

Therefore, under perpendicular `quantum' fluctuations away from the classical solutions, the toroid states which look stable in ${\cal H}_{cl}$ are also generally (meta-)stable. It
should be noted here that the fluctuation along the classical
solutions may still be flat so that the transitions from one toroid
state to another is possible. These facts justify our claims on the phase
transitions between toroid states and their existence.

As for the toroid state with the winding number one, its stability
depends on the value of $c$. The transition along the classical
solutions is almost flat so that when $c\leq \frac12$ it naturally
goes down to a whip state (Fig.\ref{fig:valley}). Therefore, it is absolutely unstable. When $c>\frac12$, one may state that it is meta-stable since attracted parts locally stabilise the state. However, the non-attracted parts are still free to move unless it gives an increase of the total energy. So, the toroid state with $N=1$ is partially stable or metastable. Its probability is given by summing over such quantum fluctuations of non-attracted parts that give the energy similar to that of the classical solution.

On the other hand, even with the attraction, the whip state is
unstable since it is not affected by the presence of the
attraction. Therefore, it would be meaningless to pick up any
particular shape of the whips and estimate its probability. Instead,
one should only estimate the probability of all the whip states that
have the similar energy ${\cal H}(a< 2\pi/L)$, by carefully counting the number of such states, or equivalently by estimating entropy. Note that, roughly speaking, the whip state is 
 more probable than a single rod state with $a=0$.

With the above reasons, it would be more appropriate to state that one of the toroid
states of $N=[x]\geq 2$ is the ground state when $l$ is much larger
than the bond length $l_b$ and $c \geq 4$ where the whip states become
negligible. Although we listed
the above reasons, we remind that
there is a first
order transition between the rod ($a=0$) and the toroid state
($N=1$). 
The potential barrier between them is given by
$\frac{2\pi^2 l}{L}$, thus the transition will be suppressed by the factor of $e^{-1}$ or smaller when
$L< 2\pi^2 l \sim 1 \mu m$ in the case of DNA with $l\sim 50nm$.

\subsection{Perturbation by the classical solutions}

In order to complete the theory at low energy, we construct the
low-energy effective Green function $G_{eff}$ from those of the toroid and the whip states, in perturbation theory. To make the function
more accessible, we fix the persistence length $l$ in what follows.
Accordingly, $G_{eff}$ becomes a function of $L$ and $W$, or
equivalently, of $c$ and $L$. In addition, $c = \frac{W}{2l}\left( \frac{L}{2\pi} \right)^2 \geq 4$ is assumed to ensure
the existence of the stable toroid states at the beginning.

Let us denote the Green function of the toroid states by $G_{T}$, and
that of the whip states by $G_{w}$. As we would like to sum over
all toroid contributions to $G_{T}$, the end-to-end vector $\vec{R}$ and
the initial and final bond vectors $(\vec{u}_i, \vec{u}_f)$ will
be omitted in $G_{T}$ as well as in $G_{w}$. Therefore, $G_{T}$ is a
function only of $c=c(\tau)$ where $c(\tau)$ is a function of the chain 
length
$\tau$ of the toroidal segment. $G_{w}$ is also a function of $\tau$:
the length of the non-attracted whipping segment in this case. Note
that, however, $G_{w}(\tau)$ does not depend on $W$ since the chain
segment is free by definition.

\begin{figure}[h]
\centering
\includegraphics[width=8cm]{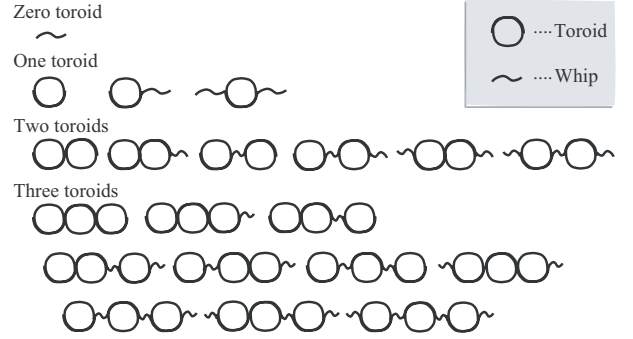}
 \caption{
Each term in eq.(\ref{pert}) is expressed by a product of the
toroids (circles) and the whips (waves). The third term is so called the
`tadpole' conformation. Below the forth term, all contributions are of
multi-tori.
}
      \label{fig:pert}
\end{figure}

With these specifications, the effective Green function can be 
constructed by the following perturbations:
\bea
&&\hspace*{-10pt}
  G_{eff}( c, L)
 =  G_{w}(L)
\nn&&
  + \Biggl[ G_{T}(c(L))
  + 2 \int_0^{L-L_{min}} \hspace{-25pt} d\tau\;\;
      G_{T}(c(L-\tau)) \, G_{w}(\tau)
  \nn&&\qquad
  + \int_{\tau_1>0, \tau_2>0, \atop \tau_1+\tau_2 < L-L_{min}}
     \hspace{-45pt} d\tau_1 d\tau_2 \;\;
     G_{w}(\tau_1) \, G_{T}(c(L-\tau_1-\tau_2)) \, G_{w}(\tau_2) \Biggr]
\nn&&
  + \Biggl[ \int_{L_{min}}^{L-L_{min}} \hspace{-25pt} d\tau\;\;
     G_{T}(c(L-\tau)) \, G_{T}(\tau)
  \nn&&\qquad
  + 2
    \int_{\tau_1>L_{min}, \tau_2>0, \atop \tau_1+\tau_2 < L-L_{min}}
     \hspace{-45pt} d\tau_1 d\tau_2 \;\;
     G_{T}(c(L-\tau_1-\tau_2)) \, G_{T}(\tau_1) \, G_{w}(\tau_2)
  \nn&&\qquad
  + \int_{\tau_1>0, \tau_2>L_{min}, \atop \tau_1+\tau_2 < L-L_{min}}
     \hspace{-45pt} d\tau_1 d\tau_2 \;\;
     G_{T}(c(L-\tau_1-\tau_2)) \, G_{w}(\tau_1) \, G_{T}(\tau_2)
  \nn&&\qquad
  + \int_{\tau_{1,3}>0, \tau_2>L_{min}, \atop \sum_{i=1}^3 \tau_i < 
L-L_{min}}
    \left[ \prod_{i=1}^3 d\tau_i \right] \;
     G_{w}(\tau_1) \times \nn
     &&\,\,\,\,\,\,\,\,\,\,\,\,\,\,\,\,\,\,\,G_{T} \left(c\left(L- \sum_{i=1}^3
\tau_i\right)\right) \, G_{T}(\tau_2) \, G_{w}(\tau_3)
  \nn&&\qquad
  + 2
   \int_{\tau_{1,3}>0, \tau_2>L_{min}, \atop \sum_{i=1}^3 \tau_i < 
L-L_{min}}
\left[ \prod_{i=1}^3 d\tau_i \right] 
     G_{T}\!\left( \!c \!\left(\! L\! -\! {\textstyle \sum_{i=1}^3 \!\tau_i}\! \right)\right) \!\times \!\nonumber\\
&&\,\,\,\,\,\,\,\,\,\,\,\,\,\,\,\,\,\,\,G_{w}(\tau_1) \, G_{T}(\tau_2) \, G_{w}(\tau_3)
  \nn&&\qquad
  + \int_{\tau_{1,2,4}>0,\tau_3>L_{min}, \atop \sum_{i=1}^4 \tau_i < 
L-L_{min}}
    \left[ \prod_{i=1}^4 d\tau_i \right] \;
     G_{w}(\tau_1) \times \,\nonumber\\ 
&&\,\,\,\,\,\,\,\,\,\,\,\,\,\,\,\,\,\,\,G_{T}\left(c\left(L- \sum_{i=1}^3
\tau_i\right)\right) \, G_{w}(\tau_2) \, G_{T}(\tau_3) \, G_{w}(\tau_4) 
\Biggr]
\nn&&
  + \cdots ,
\label{pert}
\eea
where $L_{min}$ is given by the lower bound of the conformation
parameter $c$: $c(L_{min}) = 1/2$, for the existence of a (meta-)stable
toroid state. It reads
\bea
  L_{min} = \frac{L}{\sqrt{2 c(L)}} = 2\pi \sqrt{\frac{l}{W}} .
\eea
The first bracket in eq.(\ref{pert}) gives the contributions from the
conformations which contain only one toroid, the
second bracket gives the ones with two toroids, and so on. It
should be noted that so-called `tadpole' conformation appears
in this low energy perturbation as the third term in eq.(\ref{pert}) with one toroid and one whip.
Schematically, eq.(\ref{pert}) can be depicted in Fig.\ref{fig:pert}.

Roughly speaking, in our ideal toroid with zero thickness, the
winding number $N_c$ of the dominant toroid state is proportional to
$c(\tau)$ which is a quadratic function in $\tau$. 
Besides, the minimum of the Hamiltonian is given at $a_c(N_c)$: ${\rm
Min}({\cal H} (a)) \sim - \frac{WL
N_c}{4} \sim -\frac{\pi{W^2}}{4l}\!{\left(\frac{L}{2\pi}\right)}^{\!3}$. Therefore, the ratio of
the probabilities of the toroids with different lengths $\tau$ and $L$
can be estimated by
\bea
 \frac{G_{T}(c(\tau))}{G_{T}(c(L))} \sim \frac{e^{\frac{\pi W^2}{4 l}
\left(\frac{\tau}{2\pi}\right)^3}}{e^{\frac{\pi W^2}{4 l}
\left(\frac{L}{2\pi}\right)^3}}
= e^{- \frac{W^2}{2^5 \pi^2 l} \left( L^3 - \tau^3 \right)}.
\eea
Since we assume that $L$ is large enough as $L^3 \gg \frac{2^5 \pi^2 
l}{W^2}$,
the toroid states with smaller contour length are highly suppressed
by the above factor (From $c{\geq}4$, we have $L^3 \geq {\left(2^5{\pi}^2lW\right)}^{\frac12}\left(\frac{2^5 \pi^2 l}{W^2}\right)$. If $W{\sim}O(1)$ and $l{\gg}1$, we obtain $L^3 \gg \frac{2^5 \pi^2 l}{W^2}$).

For example, this condition holds for the cases of DNA
considered in the next section. Hence, the above perturbation can be justified for a large value of $c(L)$.
Note, however, that the statistical weight for each path, or each
conformation, is not specified here, nor for the whip state and the
normalisation factor. Moreover, we have not counted other
conformations such as conventional `tadpoles' with overlaying
whips. Therefore, and unfortunately, we do not go into the precise
comparison of the above terms. Instead, we will make some remarks on
these issues in the final section.

\section{Comparison with experiments}\label{Sec_cfExperiment}

In previous sections, we have found the toroid states as the classical
solutions and found that some of them are (meta-)stable, one of which
becomes the ground state at large $c$. As mentioned, it is pointless
to ask $\vec{u}_i, \vec{u}_f$, and $\vec{R}(L)$. One of the most
physically meaningful observable is the radius of the toroid.

For large $c$, the ground state --- the dominant toroid state of the 
winding number $N_c$ can be estimated by the inequality relation (\ref{c_inequality}) of $c$:
$$
c_L^{(N_c)}< c <c_U^{(N_c)}.
$$
By its inverse relation, it reads $N_c \simeq c$, since $c_L^{(N_c)} 
\simeq N_c$ and $c_U^{(N_c)} \simeq N_c$.
Using this, we can estimate that the radius of our dominant ideal toroid behaves \bea
r_c = \frac{L}{2\pi N_c} = \frac{4\pi l}{W L}.
\eea
This result is, however, not directly applicable to the physical systems, 
because our model has the zero thickness of the chain. That is, every 
chain segment interacts equally with all the other segments accumulated 
on the same arc of the toroid.

Therefore in this section, we first introduce a finite size effect into 
our Hamiltonian. We then
estimate the mean radius of toroid and compare resulting analytical 
expression with the experiments of DNA
condensation \cite{YYK98,B96}. Also, the mean radius of the toroid
  cross section is calculated in the end.

\subsection{Finite size effect}

The finite size effect of the toroid cross section can be approximated
by the hexagonally arranged DNA chains with van der Waals type 
interactions, i.e., with the effective nearest neighbour interactions. 
Namely, if the chains are packed in a complete hexagonal cross section,
the winding numbers are $N = 7, 19, 37$, and so on.
In such cases, the number of van der Waals interactions between
segments can be counted by the links between neighbouring pairs in the
hexagonal cross section. We then obtain the number as a discrete 
function $\V_{discrete}(N)$.
We can approximate it or analytically continue to the following analytic 
function $\V(N)$:
\bea
  \V(N) = 3 N - 2\sqrt{3} \sqrt{N-\frac14}.
\eea
Thereupon, the attractive energy can be expressed by those of $N$ loops 
and the rest of the chain:
\bea
  \int_0^L\!\!\!\!ds\; V_{AT}(s)\!=\!-\frac{2\pi{W}}{a} \V(N)\!-\!W\!\!\left(\!\!L\!-\!\frac{2\pi N}{a}\! \right)\!Gap(N),
  \label{finiteVAT}
\eea
where $Gap(N)\equiv \V(N+1)-\V(N)$ is additionally introduced in order
to compensate the continuity of the potential as a function of
$a$. Note that, up to $N=3$, we need not to introduce this finite size
effect, since there is no difference between the ideal toroid and the 
hexagonally arranged case: the number of links are the same in both 
cases. Therefore, we assume $N(L)\geq 4$ for this effect.
Note also that the entanglement (knotting) effect of the chain 
arrangement is neglected.

Finally, substituting eq.(\ref{finiteVAT}) into ${\cal H}[\vec{u},W]$,
the finite size effect leads to the modified Hamiltonian:
\bea
{\cal H}(a)
 &=& \frac{Ll}{2}a^2 - \frac{2\pi W}{a} \V(N(L))\nn
& &- \frac{2\pi W}{a} \! \left(\frac{aL}{2\pi} -N(L) \right) Gap(N(L)).
\eea
Most physical observables for tightly packed toroids can be quite
accurately estimated using this Hamiltonian. For example, in principle, we can
derive the exact value of the radius of the stable toroid.
In fact, by the same analysis presented in the previous section, we 
obtain the following ``asymptotic'' relation of $N_c$ of the dominant 
toroid for large $c$:
\bea
 N_c \simeq \left( 2\sqrt{3} \, c \right)^{\frac{2}{5}} .
\label{N_c hex}
\eea

\subsection{Mapping onto experimental data}

By $r_c\equiv \frac{L}{2\pi N_c}$, we now estimate the mean radius of 
the toroid (i.e. the average of inner and outer radii) in a physical
system.
A coupling constant of (\ref{VDWhamiltonian}) can be given by
$W\!=\!\frac{1}{l_m}\!\!\left(\frac{k\epsilon}{k_BT}\right)$ where $k$
is the number of the electric dipoles in a monomer segment, each of which creates 
van der Waals interaction of the magnitude $\epsilon$. $l_m$ denotes 
the length of the monomer along the chain contour,
taken to be a half of pitch per turn (of helix) 
$l_m\simeq5\,bp=1.66\,nm$ in the end. Note we assume $l_m \sim l_b$.
Substituting $N_c$ of the dominant toroid state (\ref{N_c hex}) and the above, we obtain
\bea
r_c \simeq 
 {\left(6\pi\right)}^{-\frac15}L^{\frac15}
  {\left(\frac{l}{W}\right)}^{\!\!\frac25}
 \!\!=\!{\left(6\pi\right)}^{-\frac15}L^{\frac15}
  {\left(l_ml\right)}^{\frac25}{\left(\frac{k\epsilon}{k_BT}\right)}^{\!\!-\frac25}\!\!\!\!\!\!\!.\label{meanradiusbyVDW}
\eea
The scaling property of the first equality matches with the one in \cite{SIGPB03}. Note that a coupling constant of second nearest 
neighbour is
  vanishingly small $W_2\, {\simeq}\, 2^{-6}W=\frac{1}{64}W$, so that 
one may neglect it in this van der Waals regime.

We estimate the mean toroidal radius of T$4$ DNA in low ionic
conditions reported in \cite{YYK98}.
Using $L=57{\mu}m$, $l\,{\simeq}\,50\!\sim\!60\,nm$,
and $l_m$, the mean radius
of the toroid is
\bea
r_c = 29.09 B^{-\frac25} \sim 31.29 B^{-\frac25} {~ [nm]},
\eea
where $B\equiv \frac{k\epsilon}{k_BT}$.
This is in good agreement with the experiment $r_c \simeq 28.5$ $nm$ for $B \,{\sim}\, 1.15$. 

The same argument for the toroid formed by
Sperm DNA packaged by protamines \cite{B96} ($L=20.4{\mu}m$), gives the 
analytic value
\bea
r_c = 23.69 B^{-\frac25} \sim 25.48 B^{-\frac25} ~ [nm],
\eea
which also agrees with an experimental result $r_c {\simeq}
26.25$ $nm$ for $B \,{\sim}\, 0.85$.
Note that the former toroid is densely packed \cite{YYK98}, hence 
  hexagonal assumption could be a good approximation. It is therefore
  well expected to have the stronger attraction compared to thermal fluctuations, $B>1$. The latter has a larger diameter of the effective
segment. Thus, it may well be expected to have the weaker but large enough interaction 
with smaller $B$ to maintain a toroidal conformation.

Similarly, the mean radius of the toroid cross section can be calculated 
for the complete hexagonal cross section with a side of $n$ monomers:
\bea
r_{cross}=\frac{2+\sqrt{3}}{4}\left(n-\frac12\right)l_d \label{rcross1}
\eea
where $l_d$ is the diameter of the segment. Substituting our relation of $n$ with the winding number $N$:
$n=\frac12+\frac{1}{\sqrt3}\sqrt{N-\frac14}$, eq.(\ref{rcross1}) can be rewritten as
\bea
r_{cross}=\frac{2\sqrt{3}+3}{12}N^{\frac12}\left[1-\frac{1}{8N} + O\left(\frac{1}{N^2}\right)\right]l_d.
\eea

Using this general result, we now specify the type of interactions and
derive the expression for the mean radius of the toroid cross
section. In the case of van der Waals interactions, 
$N_c$ is given by eq.(\ref{N_c hex}), and therefore we have
\bea
r_{cross} &\simeq& \frac{3\sqrt{3}+6}{12}{\left(6\pi\right)}^{\!-\frac25}\!L^{\!\frac25}\!
  {\left(\frac{W}{l}\right)}^{\!\!\frac15}\!\!l_d \nonumber \\
 &=&\frac{3\sqrt{3}+6}{12}{\left(6\pi\right)}^{\!-\frac25}\!L^{\frac25}
  {\left(l_ml\right)}^{\!\!-\frac15}\!{\left(\frac{k\epsilon}{k_BT}\right)}^{\!\!\frac15}\!\!l_d.\label{RadiusCrosssecVDW}
\eea
Note that the scaling property $r_{cross} \sim L^{\frac25}$ is in
agreement with the one in \cite{SIGPB03} obtained in the asymptotic limit.

Similarly, we can formally consider the case of ideal toroid, although it has zero thickness.
In this case with
$N_c \simeq c$, 
we have
\bea 
r_{cross} &\simeq& \frac{2\sqrt{3}+3}{24\pi}L{\left(\frac{W}{2\,l}\right)}^{\!\!\frac12}\!l_d \nonumber \\
 &=&\frac{2\sqrt{3}+3}{24\pi}\,L\,{\left({2\,l_ml}\right)}^{\!-\!\frac12}\!{\left(\frac{k\epsilon}{k_BT}\right)}^{\!\!\frac12}\!l_d.\label{RadiusCrosssecIDEAL}
\eea

It should be noted that, with van der Waals interactions, the mean radius
of the toroid scales as $L^{\frac15}$ while that of the toroid cross section scales
as $L^{\frac25}$. In the case of ideal toroid, the mean radius
of the toroid scales as $L^{-1}$ while that of the cross section scales as $L$.

\section{Discussion and concluding remarks}

We have shown that the
(meta-)stable toroid states appear as the classical solutions of the
low energy effective theory of a semiflexible homopolymer chain ---
the nonlinear sigma model on a line segment. We have shown in this paper the complete proof of the statement that our classical solutions represent the general solution.
The novelty of the model comes from the fact that the
difficulties of the local inextensibility constraint $|\vec{u}|^2=1$ and the
delta-function potential are resolved explicitly with our
solutions in the path integral formulation (hence our theory goes further beyond 
the Gaussian approximation). Together with our
microscopic Hamiltonian, they lead us to the profound analytic curve
of the energy levels (Fig.\ref{fig:energy level}). It is also
of interests that the balance between the bending energy and the
attractive potential creates multiple local minima at $a=a_c(N)$ for
each $N$ satisfying eq.(\ref{Nbounds}). One can read off from
eq.(\ref{Nregion}) and Fig.\ref{fig:phase transition} that the number
of local minima is basically three or four in the case of our ideal
toroid. Another work on multiple local minima will be mentioned shortly.
In search of the ground state, we found that the phase
transitions occur at $c=\frac12$ and at $c=\frac{27}{16}$, and
discovered that such configurational transitions are governed by
the conformation parameter $c=\frac{W}{2l}\left( \frac{L}{2\pi}
\right)^2$. The critical point $c=\frac{27}{16}$ indicates 
the whip-toroid
or whip-dominant to toroid-dominant transition of the first
order.

In section V-A, we have shown the stability of the toroid states and
the validity of such phase transitions. We also calculated the potential barrier of the whip-toroid transition (rod-toroid transition), and indicated that, for the chains of $L< 1$ $\mu m$, the rod ($a=0$) to $N=1$ (meta-)stable toroid state transition is fairly unlikely. This would explain why it is difficult to observe short DNA toroids in experiments \cite{T05}. 
In section V-B, we have constructed the effective Green function from
those of the whip and toroid states using perturbation theory at low
energy. It naturally contains multi-tori and `tadpole' conformations. 

We finally introduced the hexagonal approximation to count the
finite size effect of the toroid cross section and established
the mapping onto the experimental data of the DNA toroid radii. Our
result is even quantitatively in good agreement with
the experiments \cite{YYK98,B96}. Hence, we conclude that our theory
is certainly an analytic theory of DNA condensation and of toroidal
condensation of many other semiflexible polymer chains with effective
van der Waals attraction, or equivalently with effective
short-range dominant attraction. Here, we presented only the
comparisons with DNA condensations, but simply by varying the parameters $l, W, L$, and $l_m$, our theory and results should fit to the same problems in similar biochemical objects.

In analogy to the classical limit $\frac{1}{\hbar} \to \infty$ in
quantum mechanics, it is assumed that the persistence length $l$ is
large enough for our low energy theory to be valid. The local
inextensibility constraint is originally given by some bond potential
such as $\lambda (|\vec{u}|-1)^2$ with the spring constant
$\lambda$. Note the low energy theory becomes invalid when the
constraint does not hold in the Hamiltonian. That is, $l$ and $\lambda$ should be sufficiently larger than $O(1)$ so that our theory remains valid. On the other hand, as $l$ approaches
$1$ or $0$, one may be able to see the transition from whip-toroid
phase to coil-globule or coil-rod phase. This is beyond our scope in
this paper, but the issue will be discussed at the end of this section.

Of particular interest on multiple local minima is the work by Kuznetsov and
Timoshenko \cite{KT99}. Using Gaussian variational method and off-lattice Monte Carlo
simulation, they numerically calculated the phase
diagram for a ring semiflexible chain in various solvent
conditions. Note that their model Hamiltonian is the same
as ours except for the harmonic spring term connecting adjacent
beads. For a given stiffness, upon increasing the
magnitude of two-body attraction, they found toroids with larger
winding numbers become more stable. This is consistent with our
findings as our conformational parameter $c$ is a function of the
magnitude of the attraction $W$. 
They also found that several distinct
toroid states (i.e. multiple minima) can exist, which are
characterized by winding numbers and are separated by first order
phase transition lines (see Fig.$4$ in \cite{KT99}). Although
they considered a ring polymer, these facts are in good agreement
with our analytic findings for an open semiflexible chain. It would be
of great interest to numerically check the existence of multiple
minima for an open semiflexible chain.

So far, we have only dealt with the whip and toroid conformations, but
there are other configurations to be explored. Numerical simulations showed that a semiflexible chain takes toroid,
collapsed rod or racquet conformations, depending on
chain length, stiffness, magnitude of interactions, temperature, and
other variables. Of particular interest is the works by Noguchi et
al. \cite{NY98,NSKY96} and Stukan et al. \cite{SIGPB03,MSIMPB05} who 
studied the dependence of stiffness on conformational properties.
They observed both toroid and collapsed rod states for
some intermediate stiffness. Upon increasing stiffness they found
toroid states are more probable.

Collapsed rods have not been present in our model. One of the reasons
is they are not classical solutions, which can only survive and become
the only candidates for the ground state at large $l$ or at small
$T$. 
Another reason may be because the inextensibility constraint $|\vec{u}|^2 = 1$ is quite strong or because collapsed rods are
energetically less favoured.
That is, our model with the constraint is simply in the quite stiff regime where the collapsed rods
are less likely. Moreover, discrete nature of the chain, which are present in most
numerical models, might allow sudden
hairpins although they are highly disfavoured in some continuum models. Indeed,
when they increase stiffness, toroid states are more probable 
\cite{NY98,MSIMPB05}. This competition in the intermediate stiffness
remains an interesting open question.

In addition to toroids and collapsed rods, tadpole like
conformations (i.e. a toroid head with long tail) have been observed
in the experiment by Noguchi et. al. \cite{NY98}. 
They also performed the Monte
Carlo simulations and found that this tadpole like structure was realised
only twice in a hundred runs. We have also dealt with a `tadpole'
conformation in the effective Green function, but it only includes the
simplest tadpole shape: a toroid with a single non-interacting
whip. Therefore, we have not counted tadpoles such as a toroid with
two whips attracting each other or a toroid with a collapsed rod or a
toroid with two collapsed rods attracting each other. To compare them,
one should first estimate the entropy of the whip and compare it to the toroids. Also, relative energy levels and statistical
weights of these conformations (toroid, collapsed rod, tadpole) including reported racquet
states of metastable intermediate \cite{MSIMPB05,SMW00,SGM02} will be studied in the forthcoming work.
Note that the whip is defined by the elongated state at low energy whose upper bound is given tentatively by $E< \frac{2\pi^2 l}{L}$. Precisely speaking, it should be the lowest bound for a chain to form a loop, which is to be explored in detail as well as the above.

As for the scaling property $r_c \sim L^{\nu}$ of the toroid radius, the exponents $\nu$ predicted in the literature
in the asymptotic limit are $\nu=\frac15$ in most
cases \cite{SIGPB03,SGM02,PW00,MKPW05}. This agrees with our precise 
asymptotic result (\ref{meanradiusbyVDW}) where both
  the parameter $c$ and the winding number $N_c$ of a dominant toroid
  are large enough. Note our model has robustness
in that it can treat chains of any finite length: a real chain is
a finite system.

However, the exponents are inconsistent with the experimentally well
known observation that the radius is independent of the chain length 
\cite{B91,B96,YYK98}.
This might suggest that the real interaction is not necessarily van der
Waals like, or at least is not a single van der Waals type interaction.
It should be noted here that combinations of our ideal toroid and its 
finite size effect can give a range of $\nu = -1 \sim \frac15$ in some 
region.

Another interesting remark is that when we apply Coulomb like interactions to our approximation, we observe the radius remains nearly constant as $L$ changes. This will be presented in the near future \cite{IK05}.

Finally, we remark that our model can be regarded as the linear sigma
model at low energy where it actually reduces to the nonlinear sigma
model. The linear sigma model is one of the most suitable models to 
describe phase transitions in quantum field theory. Although we are not formulating quantum field theory, the model actually involves a phase transition
from a constrained to non-constrained system, that is, from
constant-length bonds to spring-like bonds (Gaussian flexible
chain). This is an interesting model to be studied. Although we
have listed some questions to solve, there are obviously a lot of
problems to be investigated. Our theory could also be extended and
applied to the challenging interdisciplinary problems such as protein folding.

\vspace*{40pt}
\noindent
{\large\bf Acknowledgments}\\

We are grateful to K. Binder, V. A. Ivanov, W. Paul, and N. Uchida for their
stimulating discussions. Y.I. is grateful to K. Nagayama for his
discussions and encouragement, and to T. Araki for his
comment. N.K. is grateful to H. Noguchi, A. Cavallo and T. Kawakatsu
for stimulating discussions, and to T. A. Vilgis for his earlier
discussions of the field theory of globule-toroid transition which
led us to this direction. Y.I. acknowledges the Yukawa Institute for Theoretical Physics where this work is partially done during the YITP-W-05-04 Workshop ``Soft Matter as Structured Materials''.
N.K. acknowledges the Deutsche Forschungsgemeinschaft for financial support.

\appendix

\section{The delta function potential}

Our delta function potential expressed in the body is 
\bea
  V_{AT}(s) = -W \int_0^s ds^\prime \; \delta\left( \vec{r}(s) - \vec{r}(s^\prime) \right).\label{olddeltadef}
\eea
This is, however, rather schematic. The precise definition will be given below, sorting out two ambiguities in eq.(\ref{olddeltadef}).
One is the definition of the delta function and the other is its integration contour concerning the self-interaction contribution. The exact form of the function is given by renormalising the coupling constant, the measure, or the delta function itself appropriately, and by expecting some ultraviolet (short-range) cutoff $\epsilon > 0$ in the integration contour:
\bea
  V_{AT}(s) = 
  -W\!\!\int_0^{s-\epsilon}\!\!\!\!\!\!\![ds^\prime]_{re} \; \delta\left( \left| \vec{r}(s) - \vec{r}(s^\prime) \right| \right)\,\,\,{\rm for~} \epsilon \leq s,
\eea
otherwise it vanishes.

First, we should interpret that the delta function is not three-dimensional but one-dimensional, changing its argument from $\left( \vec{r}(s)- \vec{r}(s^\prime) \right)$ to $\left| \vec{r}(s)- \vec{r}(s^\prime) \right|$.
When the argument of the delta function has some zeros, it has the following property. Say that the argument is given by a function $g(x)$ then
\bea
  \delta\left( g(x) \right) = \sum_{x_i} \frac{\delta(x-x_i)}{|g^\prime (x_i)|}
\eea
where $\{x_i\}$ is the roots of $g(x)$. Accordingly, at around every zero,
the integration over $x$ gives a $|g^\prime(x_i)|^{-1}$
contribution. By definition, our potential should not have such a
contribution, so that the amplitude must be normalised
appropriately. For example, such a renormalisation of the measure can
be achieved by
\bea
  \int_{-\infty}^{\infty} [dx]_{re}\; \delta\left( g(x) \right)
  &=& 
  \int_{-\infty}^{\infty} [dx]_{re}\; \sum_{x_i} \frac{\delta(x-x_i)}{|g^\prime (x_i)|}
\nn
  &=& 
  \sum_{x_i} \int_{x_i} dx |g^\prime (x_i)|\; \frac{\delta(x-x_i)}{|g^\prime (x_i)|}
\nn
  &=& 
  \sum_{x_i} \int_{x_i} dx\; \delta(x-x_i)
\nn
  &=&
  {\rm Number ~of ~zeros} .
\eea
As given in the second line of the above expression, the measure
is renormalised such that it cancels all the denominators of the delta
function potential. 

Another example is to renormalise the delta function as follows:
\bea
  \int_0^s ds^\prime\; \delta^{re} ( g(s^\prime) )
  &\equiv& \int_0^s ds^\prime\; \frac{dg(s^\prime)}{ds^\prime} \delta( g(s^\prime) )\nn
  &=& \int_C d g(s^\prime) \delta( g(s^\prime) ) , 
\eea
where $C$ is given by the one-parameter contour of $g(s^\prime)$ from
$s^\prime = 0$ to $s^\prime= s$. 
Therefore, we implicitly include one of these renormalisations in eq.(\ref{olddeltadef}).

The second ambiguity is that, when the integration contour is from $0$
to $s$, it arises a self-interaction between a point at $s$ and an adjacent point at $(s - \epsilon)$ with some infinitesimal positive parameter $\epsilon$.
In order to avoid such a self-interaction, we have implicitly introduced an ultraviolet cutoff $\epsilon$ ($\epsilon>0$) in the form of $V_{AT}(s)$:
\bea
  \int_0^{s-\epsilon} ds^\prime \delta( \cdots ).
\eea
Note that $V_{AT}(s < \epsilon)$ is defined as nil.

\section{SO(3) transformation}
\label{sec:SO(3)}

The dimension of the generators of $SO(3)$ are three: $T^i$ for $i=1$
to $3$ and global $SO(3)$ transformation can be given by its
exponential mapping: $e^{g_i T^i}$ where $g_i$ are some
arbitrary parameters. The matrix form of the generators on the
fundamental representations are antisymmetric $\{ \left( T^i \right)_{jk}\}$
where $j$ and $k$ are matrix suffices running from 1 to 3. Its adjoint
representation is given by $(T^i)_{jk} = \epsilon_{ijk}$ where
  $\epsilon_{ijk}$ is the complete antisymmetric tensor.

The $SO(3)$ infinitesimal transformation of the bond vector $\vec{u}$ is given by 
\bea
 \delta u_j = g_i T^i_{jk} u_k, 
\eea
where $i,k$ are summed over and $g_i$ are infinitesimal parameters in this case.
Clearly, the action is invariant under such transformations:
\bea
  (u^i)^\prime (u_i)^\prime - u^i u_i
  &=& (u^i + \delta u^i) (u_i + \delta u_i) - |u|^2\nn
  &=& g_k \left( (u^j T_{ji}^{k}) u_i + u^i (T_{ij}^{k} u_j) \right) 
  \nn
  &=& g_k \left( (u_i T^k u_j)^{T} + (u_i T^k u_j) \right)
  \nn
  &=& 0,
\eea
since $T^k$ is antisymmetric $(T^k)^T = -(T^k)$. Note that $u_i$ and $\pa u_i$ have the same property under $SO(3)$. For simplicity, let
us take its adjoint representation $T^i_{jk}= \epsilon_{ijk}$ and
write down the transformation law in the polar coordinates 
\bea
 \delta u_j = g_i \epsilon_{ijk} u_k,
\eea
where $i, k$ are summed over. Substituting the polar decomposition
with the constraint $|\vec{u}|^2 = 1$, one obtains
\bea
  \delta u_x 
  &=& \delta(\sin \theta_u \cos \phi_u) 
  = \left( g_3 \sin \theta_u \sin \phi_u - g_2 \cos \theta_u \right) 
  \nn&& \Rightarrow
  \delta \theta_u \cos \theta_u \cos \phi_u - \delta \phi_u \sin \theta_u \sin \phi_u\nn
  &&\,\,\,\,\,\,\,\,\,= \left( g_3 \sin \theta_u \sin \phi_u - g_2 \cos \theta_u \right) ,
  \nn
  \delta u_y
  &=& \delta( \sin \theta_u \sin \phi_u ) 
  = \left( g_1 \cos \theta_u - g_3 \sin \theta_u \cos \phi_u \right) 
  \nn&& \Rightarrow
  \delta \theta_u \cos \theta_u \sin \phi_u + \delta \phi_u \sin \theta_u \cos \phi_u\nn
  &&\,\,\,\,\,\,\,\,\,= \left( g_1 \cos \theta_u - g_3 \sin \theta_u \cos \phi_u \right) ,
  \nn
  \delta u_z
  &=& \delta( \cos \theta_u ) 
  = \sin \theta_u \left( g_2 \cos \phi_u - g_1 \sin \phi_u \right)
  \nn&& \Rightarrow
  \delta \theta_u = \left( g_1 \sin \phi_u - g_2 \cos \phi_u \right) .
\eea
Thus, the transformations of $\theta_u$ is shown in the last line.
From the first and second transformations, one finds
\bea
&&  (second) \cos \phi_u - (first) \sin \phi_u 
\nn&& \Rightarrow 
  \delta \phi_u \sin \theta_u 
  = \left( g_1 \cos \theta_u - g_3 \sin \theta_u \cos \phi_u \right) \cos \phi_u\nn
  &&\,\,\,\,\,\,\,\,\,\,\,\,\,\,\,\,\,\,\,\,\,\,\,\,\,\,\,\,\,\,\,\,\,\,\,\,- \left( g_3 \sin \theta_u \sin \phi_u - g_2 \cos \theta_u \right) \sin \phi_u
  \nn&&\quad
  \delta \phi_u 
  = \left( \cot \theta_u \left( g_1 \cos \phi_u + g_2 \sin \phi_u  \right) - g_3 \right).
\eea
Therefore, the $SO(3)$ transformations are given by
\bea
  \delta \theta_u &=& g_1 \sin \phi_u - g_2 \cos \phi_u ,
  \nn
  \delta \phi_u 
  &=& \cot \theta_u \left( g_1 \cos \phi_u + g_2 \sin \phi_u  \right) - g_3 ,
\eea
where $g_i$ are arbitrary infinitesimal parameters which represent rotations around $i$-axis, {\it i.e.}, $x$-, $y$-, and $z$-axes.



\begin{thebibliography}{52}
\expandafter\ifx\csname natexlab\endcsname\relax\def\natexlab#1{#1}\fi
\expandafter\ifx\csname bibnamefont\endcsname\relax
  \def\bibnamefont#1{#1}\fi
\expandafter\ifx\csname bibfnamefont\endcsname\relax
  \def\bibfnamefont#1{#1}\fi
\expandafter\ifx\csname citenamefont\endcsname\relax
  \def\citenamefont#1{#1}\fi
\expandafter\ifx\csname url\endcsname\relax
  \def\url#1{\texttt{#1}}\fi
\expandafter\ifx\csname urlprefix\endcsname\relax\def\urlprefix{URL }\fi
\providecommand{\bibinfo}[2]{#2}
\providecommand{\eprint}[2][]{\url{#2}}

\bibitem[{\citenamefont{Doi and Edwards}(1986)}]{DE86}
\bibinfo{author}{\bibfnamefont{M.}~\bibnamefont{Doi}} \bibnamefont{and}
  \bibinfo{author}{\bibfnamefont{S.~F.} \bibnamefont{Edwards}},
  \emph{\bibinfo{title}{The theory of polymer dynamics}}
  (\bibinfo{publisher}{Clarendon Press}, \bibinfo{address}{Oxford},
  \bibinfo{year}{1986}).

\bibitem[{\citenamefont{de~Gennes}(1979)}]{DG79}
\bibinfo{author}{\bibfnamefont{P.~G.} \bibnamefont{de~Gennes}},
  \emph{\bibinfo{title}{Scaling Concepts in Polymer Physics}}
  (\bibinfo{publisher}{Cornell University Press}, \bibinfo{address}{New York},
  \bibinfo{year}{1979}).

\bibitem[{\citenamefont{Grosberg and Khokhlov}(1994)}]{GK94}
\bibinfo{author}{\bibfnamefont{A.~Y.} \bibnamefont{Grosberg}} \bibnamefont{and}
  \bibinfo{author}{\bibfnamefont{A.~R.} \bibnamefont{Khokhlov}},
  \emph{\bibinfo{title}{Statistical physics of macromolecules}}
  (\bibinfo{publisher}{American Institute of Physics}, \bibinfo{address}{New
  York}, \bibinfo{year}{1994}).

\bibitem[{\citenamefont{Finkelstein and Ptitsyn}(2002)}]{FP02}
\bibinfo{author}{\bibfnamefont{A.~V.} \bibnamefont{Finkelstein}}
  \bibnamefont{and} \bibinfo{author}{\bibfnamefont{O.}~\bibnamefont{Ptitsyn}},
  \emph{\bibinfo{title}{Protein Physics : A Course of Lectures}}
  (\bibinfo{publisher}{Academic Press}, \bibinfo{address}{London},
  \bibinfo{year}{2002}).

\bibitem[{\citenamefont{Perkins et~al.}(1995)\citenamefont{Perkins, Smith,
  Larson, and Chu}}]{PSLC95}
\bibinfo{author}{\bibfnamefont{T.~T.} \bibnamefont{Perkins}},
  \bibinfo{author}{\bibfnamefont{D.~E.} \bibnamefont{Smith}},
  \bibinfo{author}{\bibfnamefont{R.~G.} \bibnamefont{Larson}},
  \bibnamefont{and} \bibinfo{author}{\bibfnamefont{S.}~\bibnamefont{Chu}},
  \bibinfo{journal}{Science} \textbf{\bibinfo{volume}{268}},
  \bibinfo{pages}{83} (\bibinfo{year}{1995}).

\bibitem[{\citenamefont{Strick et~al.}(1996)\citenamefont{Strick, Allemand,
  Bensimon, Bensimon, and Croquette}}]{SABBC96}
\bibinfo{author}{\bibfnamefont{T.~R.} \bibnamefont{Strick}},
  \bibinfo{author}{\bibfnamefont{J.~F.} \bibnamefont{Allemand}},
  \bibinfo{author}{\bibfnamefont{D.}~\bibnamefont{Bensimon}},
  \bibinfo{author}{\bibfnamefont{A.}~\bibnamefont{Bensimon}}, \bibnamefont{and}
  \bibinfo{author}{\bibfnamefont{V.}~\bibnamefont{Croquette}},
  \bibinfo{journal}{Science} \textbf{\bibinfo{volume}{271}},
  \bibinfo{pages}{1835} (\bibinfo{year}{1996}).

\bibitem[{\citenamefont{Perkins et~al.}(1997)\citenamefont{Perkins, Smith, and
  Chu}}]{PSC97}
\bibinfo{author}{\bibfnamefont{T.~T.} \bibnamefont{Perkins}},
  \bibinfo{author}{\bibfnamefont{D.~E.} \bibnamefont{Smith}}, \bibnamefont{and}
  \bibinfo{author}{\bibfnamefont{S.}~\bibnamefont{Chu}},
  \bibinfo{journal}{Science} \textbf{\bibinfo{volume}{276}},
  \bibinfo{pages}{2016} (\bibinfo{year}{1997}).

\bibitem[{\citenamefont{Goff et~al.}(2002)\citenamefont{Goff, Hallatschek,
  Frey, and Amblard}}]{GHFA02}
\bibinfo{author}{\bibfnamefont{L.~L.} \bibnamefont{Goff}},
  \bibinfo{author}{\bibfnamefont{O.}~\bibnamefont{Hallatschek}},
  \bibinfo{author}{\bibfnamefont{E.}~\bibnamefont{Frey}}, \bibnamefont{and}
  \bibinfo{author}{\bibfnamefont{F.}~\bibnamefont{Amblard}},
  \bibinfo{journal}{Phys.\ Rev.\ Lett.} \textbf{\bibinfo{volume}{89}},
  \bibinfo{pages}{258101} (\bibinfo{year}{2002}).

\bibitem[{\citenamefont{Smith et~al.}(2001)\citenamefont{Smith, Tans, Smith,
  Grimes, Anderson, and Bustamante}}]{STSGAB01}
\bibinfo{author}{\bibfnamefont{D.~E.} \bibnamefont{Smith}},
  \bibinfo{author}{\bibfnamefont{S.~J.} \bibnamefont{Tans}},
  \bibinfo{author}{\bibfnamefont{S.~B.} \bibnamefont{Smith}},
  \bibinfo{author}{\bibfnamefont{S.}~\bibnamefont{Grimes}},
  \bibinfo{author}{\bibfnamefont{D.~L.} \bibnamefont{Anderson}},
  \bibnamefont{and}
  \bibinfo{author}{\bibfnamefont{C.}~\bibnamefont{Bustamante}},
  \bibinfo{journal}{Nature} \textbf{\bibinfo{volume}{413}},
  \bibinfo{pages}{748} (\bibinfo{year}{2001}).

\bibitem[{\citenamefont{Gosule and Schellman}(1976)}]{GS76}
\bibinfo{author}{\bibfnamefont{L.~C.} \bibnamefont{Gosule}} \bibnamefont{and}
  \bibinfo{author}{\bibfnamefont{J.~A.} \bibnamefont{Schellman}},
  \bibinfo{journal}{Nature} \textbf{\bibinfo{volume}{259}},
  \bibinfo{pages}{333} (\bibinfo{year}{1976}).

\bibitem[{\citenamefont{Bloomfield}(1991)}]{B91}
\bibinfo{author}{\bibfnamefont{V.~A.} \bibnamefont{Bloomfield}},
  \bibinfo{journal}{Biopolymers} \textbf{\bibinfo{volume}{31}},
  \bibinfo{pages}{1471} (\bibinfo{year}{1991}).

\bibitem[{\citenamefont{Bloomfield}(1996)}]{B96}
\bibinfo{author}{\bibfnamefont{V.~A.} \bibnamefont{Bloomfield}},
  \bibinfo{journal}{Curr.\ Opinion\ Struct.\ Biol.}
  \textbf{\bibinfo{volume}{6}}, \bibinfo{pages}{334} (\bibinfo{year}{1996}).

\bibitem[{\citenamefont{Yoshikawa et~al.}(1999)\citenamefont{Yoshikawa,
  Yoshikawa, and Kanbe}}]{YYK98}
\bibinfo{author}{\bibfnamefont{Y.}~\bibnamefont{Yoshikawa}},
  \bibinfo{author}{\bibfnamefont{K.}~\bibnamefont{Yoshikawa}},
  \bibnamefont{and} \bibinfo{author}{\bibfnamefont{T.}~\bibnamefont{Kanbe}},
  \bibinfo{journal}{Langmuir} \textbf{\bibinfo{volume}{15}},
  \bibinfo{pages}{4085} (\bibinfo{year}{1999}).

\bibitem[{\citenamefont{Conwell et~al.}(2003)\citenamefont{Conwell, Vilfan, and
  Hud}}]{CVH03}
\bibinfo{author}{\bibfnamefont{C.~C.} \bibnamefont{Conwell}},
  \bibinfo{author}{\bibfnamefont{I.~D.} \bibnamefont{Vilfan}},
  \bibnamefont{and} \bibinfo{author}{\bibfnamefont{N.~V.} \bibnamefont{Hud}},
  \bibinfo{journal}{Proc.\ Natl.\ Acad.\ Sci.} \textbf{\bibinfo{volume}{100}},
  \bibinfo{pages}{9296} (\bibinfo{year}{2003}).

\bibitem[{\citenamefont{Hud and Downing}(2001)}]{HD01}
\bibinfo{author}{\bibfnamefont{N.~V.} \bibnamefont{Hud}} \bibnamefont{and}
  \bibinfo{author}{\bibfnamefont{K.~H.} \bibnamefont{Downing}},
  \bibinfo{journal}{Proc.\ Natl.\ Acad.\ Sci.} \textbf{\bibinfo{volume}{98}},
  \bibinfo{pages}{14925} (\bibinfo{year}{2001}).

\bibitem[{\citenamefont{Lifshitz et~al.}(1978)\citenamefont{Lifshitz, Grosberg,
  and Khokhlov}}]{LGK78}
\bibinfo{author}{\bibfnamefont{I.~M.} \bibnamefont{Lifshitz}},
  \bibinfo{author}{\bibfnamefont{A.~Y.} \bibnamefont{Grosberg}},
  \bibnamefont{and} \bibinfo{author}{\bibfnamefont{A.~R.}
  \bibnamefont{Khokhlov}}, \bibinfo{journal}{Rev.\ Mod.\ Phys.}
  \textbf{\bibinfo{volume}{50}}, \bibinfo{pages}{683} (\bibinfo{year}{1978}).

\bibitem[{\citenamefont{Kholodenko and Freed}(1984)}]{KF84}
\bibinfo{author}{\bibfnamefont{A.~L.} \bibnamefont{Kholodenko}}
  \bibnamefont{and} \bibinfo{author}{\bibfnamefont{K.~F.} \bibnamefont{Freed}},
  \bibinfo{journal}{J.\ Phys.\ A:\ Math.\ Gen.} \textbf{\bibinfo{volume}{17}},
  \bibinfo{pages}{2703} (\bibinfo{year}{1984}).

\bibitem[{\citenamefont{de~Gennes}(1985)}]{DG85}
\bibinfo{author}{\bibfnamefont{P.~G.} \bibnamefont{de~Gennes}},
  \bibinfo{journal}{J.\ Phys.\ (France)\ Lett.} \textbf{\bibinfo{volume}{46}},
  \bibinfo{pages}{L639} (\bibinfo{year}{1985}).

\bibitem[{\citenamefont{Kuznetsov
  et~al.}(1996{\natexlab{a}})\citenamefont{Kuznetsov, Timoshenko, and
  Dawson}}]{KTD96a}
\bibinfo{author}{\bibfnamefont{Y.~A.} \bibnamefont{Kuznetsov}},
  \bibinfo{author}{\bibfnamefont{E.~G.} \bibnamefont{Timoshenko}},
  \bibnamefont{and} \bibinfo{author}{\bibfnamefont{K.~A.}
  \bibnamefont{Dawson}}, \bibinfo{journal}{J.\ Chem.\ Phys.}
  \textbf{\bibinfo{volume}{104}}, \bibinfo{pages}{3338}
  (\bibinfo{year}{1996}{\natexlab{a}}).

\bibitem[{\citenamefont{Abrams et~al.}(2002)\citenamefont{Abrams, Lee, and
  Obukhov}}]{ALO02}
\bibinfo{author}{\bibfnamefont{C.~F.} \bibnamefont{Abrams}},
  \bibinfo{author}{\bibfnamefont{N.}~\bibnamefont{Lee}}, \bibnamefont{and}
  \bibinfo{author}{\bibfnamefont{S.}~\bibnamefont{Obukhov}},
  \bibinfo{journal}{Europhys.\ Lett.} \textbf{\bibinfo{volume}{59}},
  \bibinfo{pages}{391} (\bibinfo{year}{2002}).

\bibitem[{\citenamefont{Kikuchi et~al.}(2005)\citenamefont{Kikuchi, Ryder,
  Pooley, and Yeomans}}]{KRPY05}
\bibinfo{author}{\bibfnamefont{N.}~\bibnamefont{Kikuchi}},
  \bibinfo{author}{\bibfnamefont{J.~F.} \bibnamefont{Ryder}},
  \bibinfo{author}{\bibfnamefont{C.~M.} \bibnamefont{Pooley}},
  \bibnamefont{and} \bibinfo{author}{\bibfnamefont{J.~M.}
  \bibnamefont{Yeomans}}, \bibinfo{journal}{Phys.\ Rev.\ E}
  \textbf{\bibinfo{volume}{71}}, \bibinfo{pages}{061804}
  (\bibinfo{year}{2005}).

\bibitem[{\citenamefont{Kleinert}(2004)}]{K04}
\bibinfo{author}{\bibfnamefont{H.}~\bibnamefont{Kleinert}},
  \emph{\bibinfo{title}{Path Integrals in Quantum Mechanics, Statistics, and
  Polymer Physics, and Financial Markets}} (\bibinfo{publisher}{World
  Scientific Publishing Company}, \bibinfo{year}{2004}).

\bibitem[{\citenamefont{Freed}(1972)}]{KF72}
\bibinfo{author}{\bibfnamefont{K.~F.} \bibnamefont{Freed}},
  \bibinfo{journal}{Adv.\ Chem.\ Phys.} \textbf{\bibinfo{volume}{22}},
  \bibinfo{pages}{1} (\bibinfo{year}{1972}).

\bibitem[{\citenamefont{Grosberg and Khokhlov}(1981)}]{GK81}
\bibinfo{author}{\bibfnamefont{A.~Y.} \bibnamefont{Grosberg}} \bibnamefont{and}
  \bibinfo{author}{\bibfnamefont{A.~R.} \bibnamefont{Khokhlov}},
  \bibinfo{journal}{Adv.\ Polym.\ Sci.} \textbf{\bibinfo{volume}{41}},
  \bibinfo{pages}{53} (\bibinfo{year}{1981}).

\bibitem[{\citenamefont{Stukan et~al.}(2003)\citenamefont{Stukan, Ivanov,
  Grosberg, Paul, and Binder}}]{SIGPB03}
\bibinfo{author}{\bibfnamefont{M.~R.} \bibnamefont{Stukan}},
  \bibinfo{author}{\bibfnamefont{V.~A.} \bibnamefont{Ivanov}},
  \bibinfo{author}{\bibfnamefont{A.~Y.} \bibnamefont{Grosberg}},
  \bibinfo{author}{\bibfnamefont{W.}~\bibnamefont{Paul}}, \bibnamefont{and}
  \bibinfo{author}{\bibfnamefont{K.}~\bibnamefont{Binder}},
  \bibinfo{journal}{J.\ Chem.\ Phys.} \textbf{\bibinfo{volume}{118}},
  \bibinfo{pages}{3392} (\bibinfo{year}{2003}).

\bibitem[{\citenamefont{Hud et~al.}(1995)\citenamefont{Hud, Downing, and
  Balhorn}}]{HDB95}
\bibinfo{author}{\bibfnamefont{N.~V.} \bibnamefont{Hud}},
  \bibinfo{author}{\bibfnamefont{K.~H.} \bibnamefont{Downing}},
  \bibnamefont{and} \bibinfo{author}{\bibfnamefont{R.}~\bibnamefont{Balhorn}},
  \bibinfo{journal}{Proc.\ Natl.\ Acad.\ Sci.} \textbf{\bibinfo{volume}{92}},
  \bibinfo{pages}{3581} (\bibinfo{year}{1995}).

\bibitem[{\citenamefont{Ubbink and Odijk}(1996)}]{UO96}
\bibinfo{author}{\bibfnamefont{J.}~\bibnamefont{Ubbink}} \bibnamefont{and}
  \bibinfo{author}{\bibfnamefont{T.}~\bibnamefont{Odijk}},
  \bibinfo{journal}{Europhys.\ Lett.} \textbf{\bibinfo{volume}{33}},
  \bibinfo{pages}{353} (\bibinfo{year}{1996}).

\bibitem[{\citenamefont{Schnurr et~al.}(2000)\citenamefont{Schnurr, MacKintosh,
  and Williams}}]{SMW00}
\bibinfo{author}{\bibfnamefont{B.}~\bibnamefont{Schnurr}},
  \bibinfo{author}{\bibfnamefont{F.~C.} \bibnamefont{MacKintosh}},
  \bibnamefont{and} \bibinfo{author}{\bibfnamefont{D.~R.~M.}
  \bibnamefont{Williams}}, \bibinfo{journal}{Europhys.\ Lett.}
  \textbf{\bibinfo{volume}{51}}, \bibinfo{pages}{279} (\bibinfo{year}{2000}).

\bibitem[{\citenamefont{Schnurr et~al.}(2002)\citenamefont{Schnurr, Gittes, and
  MacKintosh}}]{SGM02}
\bibinfo{author}{\bibfnamefont{B.}~\bibnamefont{Schnurr}},
  \bibinfo{author}{\bibfnamefont{F.}~\bibnamefont{Gittes}}, \bibnamefont{and}
  \bibinfo{author}{\bibfnamefont{F.~C.} \bibnamefont{MacKintosh}},
  \bibinfo{journal}{Phys.\ Rev.\ E} \textbf{\bibinfo{volume}{65}},
  \bibinfo{pages}{061904} (\bibinfo{year}{2002}).

\bibitem[{\citenamefont{Pereira and Williams}(2000)}]{PW00}
\bibinfo{author}{\bibfnamefont{G.~G.} \bibnamefont{Pereira}} \bibnamefont{and}
  \bibinfo{author}{\bibfnamefont{D.~R.~M.} \bibnamefont{Williams}},
  \bibinfo{journal}{Europhys.\ Lett.} \textbf{\bibinfo{volume}{50}},
  \bibinfo{pages}{559} (\bibinfo{year}{2000}).

\bibitem[{\citenamefont{Miller et~al.}(2005)\citenamefont{Miller, Keentok,
  Pereira, and Williams}}]{MKPW05}
\bibinfo{author}{\bibfnamefont{I.~C.~B.} \bibnamefont{Miller}},
  \bibinfo{author}{\bibfnamefont{M.}~\bibnamefont{Keentok}},
  \bibinfo{author}{\bibfnamefont{G.~G.} \bibnamefont{Pereira}},
  \bibnamefont{and} \bibinfo{author}{\bibfnamefont{D.~R.~M.}
  \bibnamefont{Williams}}, \bibinfo{journal}{Phys.\ Rev.\ E}
  \textbf{\bibinfo{volume}{71}}, \bibinfo{pages}{031802}
  (\bibinfo{year}{2005}).

\bibitem[{\citenamefont{Park et~al.}(1998)\citenamefont{Park, Harries, and
  Gelbart}}]{PHG98}
\bibinfo{author}{\bibfnamefont{S.~Y.} \bibnamefont{Park}},
  \bibinfo{author}{\bibfnamefont{D.}~\bibnamefont{Harries}}, \bibnamefont{and}
  \bibinfo{author}{\bibfnamefont{W.~M.} \bibnamefont{Gelbart}},
  \bibinfo{journal}{Biophys.\ J.} \textbf{\bibinfo{volume}{75}},
  \bibinfo{pages}{714} (\bibinfo{year}{1998}).

\bibitem[{\citenamefont{Takenaka et~al.}(2005)\citenamefont{Takenaka,
  Yoshikawa, Yoshikawa, Koyama, and Kanbe}}]{TYYKK05}
\bibinfo{author}{\bibfnamefont{Y.}~\bibnamefont{Takenaka}},
  \bibinfo{author}{\bibfnamefont{K.}~\bibnamefont{Yoshikawa}},
  \bibinfo{author}{\bibfnamefont{Y.}~\bibnamefont{Yoshikawa}},
  \bibinfo{author}{\bibfnamefont{Y.}~\bibnamefont{Koyama}}, \bibnamefont{and}
  \bibinfo{author}{\bibfnamefont{T.}~\bibnamefont{Kanbe}},
  \bibinfo{journal}{J.\ Chem.\ Phys.} \textbf{\bibinfo{volume}{123}},
  \bibinfo{pages}{014902} (\bibinfo{year}{2005}).

\bibitem[{\citenamefont{Montesi et~al.}(2004)\citenamefont{Montesi, Pasquali,
  and MacKintosh}}]{MPM04}
\bibinfo{author}{\bibfnamefont{A.}~\bibnamefont{Montesi}},
  \bibinfo{author}{\bibfnamefont{M.}~\bibnamefont{Pasquali}}, \bibnamefont{and}
  \bibinfo{author}{\bibfnamefont{F.~C.} \bibnamefont{MacKintosh}},
  \bibinfo{journal}{Phys.\ Rev.\ E} \textbf{\bibinfo{volume}{69}},
  \bibinfo{pages}{021916} (\bibinfo{year}{2004}).

\bibitem[{\citenamefont{Cooke and Williams}(2004)}]{CW04}
\bibinfo{author}{\bibfnamefont{I.~R.} \bibnamefont{Cooke}} \bibnamefont{and}
  \bibinfo{author}{\bibfnamefont{D.~R.~M.} \bibnamefont{Williams}},
  \bibinfo{journal}{Physica A} \textbf{\bibinfo{volume}{339}},
  \bibinfo{pages}{45} (\bibinfo{year}{2004}).

\bibitem[{\citenamefont{Noguchi et~al.}(1996)\citenamefont{Noguchi, Saito,
  Kidoaki, and Yoshikawa}}]{NSKY96}
\bibinfo{author}{\bibfnamefont{H.}~\bibnamefont{Noguchi}},
  \bibinfo{author}{\bibfnamefont{S.}~\bibnamefont{Saito}},
  \bibinfo{author}{\bibfnamefont{S.}~\bibnamefont{Kidoaki}}, \bibnamefont{and}
  \bibinfo{author}{\bibfnamefont{K.}~\bibnamefont{Yoshikawa}},
  \bibinfo{journal}{Chem.\ Phys.\ Lett.} \textbf{\bibinfo{volume}{261}},
  \bibinfo{pages}{527} (\bibinfo{year}{1996}).

\bibitem[{\citenamefont{Noguchi and Yoshikawa}(1998)}]{NY98}
\bibinfo{author}{\bibfnamefont{H.}~\bibnamefont{Noguchi}} \bibnamefont{and}
  \bibinfo{author}{\bibfnamefont{K.}~\bibnamefont{Yoshikawa}},
  \bibinfo{journal}{J.\ Chem.\ Phys.} \textbf{\bibinfo{volume}{109}},
  \bibinfo{pages}{5070} (\bibinfo{year}{1998}).

\bibitem[{\citenamefont{Kuznetsov and Timoshenko}(1999)}]{KT99}
\bibinfo{author}{\bibfnamefont{Y.~A.} \bibnamefont{Kuznetsov}}
  \bibnamefont{and} \bibinfo{author}{\bibfnamefont{E.~G.}
  \bibnamefont{Timoshenko}}, \bibinfo{journal}{J.\ Chem.\ Phys.}
  \textbf{\bibinfo{volume}{111}}, \bibinfo{pages}{3744} (\bibinfo{year}{1999}).

\bibitem[{\citenamefont{Kuznetsov
  et~al.}(1996{\natexlab{b}})\citenamefont{Kuznetsov, Timoshenko, and
  Dawson}}]{KTD96}
\bibinfo{author}{\bibfnamefont{Y.~A.} \bibnamefont{Kuznetsov}},
  \bibinfo{author}{\bibfnamefont{E.~G.} \bibnamefont{Timoshenko}},
  \bibnamefont{and} \bibinfo{author}{\bibfnamefont{K.~A.}
  \bibnamefont{Dawson}}, \bibinfo{journal}{J.\ Chem.\ Phys.}
  \textbf{\bibinfo{volume}{105}}, \bibinfo{pages}{7116}
  (\bibinfo{year}{1996}{\natexlab{b}}).

\bibitem[{\citenamefont{Ivanov et~al.}(1998)\citenamefont{Ivanov, Paul, and
  Binder}}]{IPB98}
\bibinfo{author}{\bibfnamefont{V.~A.} \bibnamefont{Ivanov}},
  \bibinfo{author}{\bibfnamefont{W.}~\bibnamefont{Paul}}, \bibnamefont{and}
  \bibinfo{author}{\bibfnamefont{K.}~\bibnamefont{Binder}},
  \bibinfo{journal}{J.\ Chem.\ Phys.} \textbf{\bibinfo{volume}{109}},
  \bibinfo{pages}{5659} (\bibinfo{year}{1998}).

\bibitem[{\citenamefont{Martemyanova et~al.}(2005)\citenamefont{Martemyanova,
  Stukan, Ivanov, Mueller, Paul, and Binder}}]{MSIMPB05}
\bibinfo{author}{\bibfnamefont{J.~A.} \bibnamefont{Martemyanova}},
  \bibinfo{author}{\bibfnamefont{M.~R.} \bibnamefont{Stukan}},
  \bibinfo{author}{\bibfnamefont{V.~A.} \bibnamefont{Ivanov}},
  \bibinfo{author}{\bibfnamefont{M.}~\bibnamefont{Mueller}},
  \bibinfo{author}{\bibfnamefont{W.}~\bibnamefont{Paul}}, \bibnamefont{and}
  \bibinfo{author}{\bibfnamefont{K.}~\bibnamefont{Binder}},
  \bibinfo{journal}{J.\ Chem.\ Phys.} \textbf{\bibinfo{volume}{122}},
  \bibinfo{pages}{174907} (\bibinfo{year}{2005}).

\bibitem[{\citenamefont{Eickbush and Moudrianakis}(1978)}]{EM78}
\bibinfo{author}{\bibfnamefont{T.~H.} \bibnamefont{Eickbush}} \bibnamefont{and}
  \bibinfo{author}{\bibfnamefont{E.~N.} \bibnamefont{Moudrianakis}},
  \bibinfo{journal}{Cell} \textbf{\bibinfo{volume}{13}}, \bibinfo{pages}{295}
  (\bibinfo{year}{1978}).

\bibitem[{\citenamefont{Plum et~al.}(1990)\citenamefont{Plum, Arscott, and
  Bloomfield}}]{PAB90}
\bibinfo{author}{\bibfnamefont{G.~E.} \bibnamefont{Plum}},
  \bibinfo{author}{\bibfnamefont{P.~G.} \bibnamefont{Arscott}},
  \bibnamefont{and} \bibinfo{author}{\bibfnamefont{V.~A.}
  \bibnamefont{Bloomfield}}, \bibinfo{journal}{Biopolymers}
  \textbf{\bibinfo{volume}{30}}, \bibinfo{pages}{631} (\bibinfo{year}{1990}).

\bibitem[{\citenamefont{Fang and Hoh}(1999)}]{FH99}
\bibinfo{author}{\bibfnamefont{Y.}~\bibnamefont{Fang}} \bibnamefont{and}
  \bibinfo{author}{\bibfnamefont{J.~H.} \bibnamefont{Hoh}},
  \bibinfo{journal}{FEBS\ Lett.} \textbf{\bibinfo{volume}{459}},
  \bibinfo{pages}{173} (\bibinfo{year}{1999}).

\bibitem[{\citenamefont{Bottcher}(1998)}]{B98}
\bibinfo{author}{\bibfnamefont{C.}~\bibnamefont{Bottcher}},
  \bibinfo{journal}{J.\ Am.\ Chem.\ Soc.} \textbf{\bibinfo{volume}{120}},
  \bibinfo{pages}{12} (\bibinfo{year}{1998}).

\bibitem[{\citenamefont{Hamprecht and Kleinert}(2005)}]{HK05}
\bibinfo{author}{\bibfnamefont{B.}~\bibnamefont{Hamprecht}} \bibnamefont{and}
  \bibinfo{author}{\bibfnamefont{H.}~\bibnamefont{Kleinert}},
  \bibinfo{journal}{Phys.\ Rev.\ E} \textbf{\bibinfo{volume}{71}},
  \bibinfo{pages}{031803} (\bibinfo{year}{2005}).

\bibitem[{\citenamefont{Spakowitz and Wang}(2003)}]{SW03}
\bibinfo{author}{\bibfnamefont{A.~J.} \bibnamefont{Spakowitz}}
  \bibnamefont{and} \bibinfo{author}{\bibfnamefont{Z.~G.} \bibnamefont{Wang}},
  \bibinfo{journal}{Phys.\ Rev.\ Lett.} \textbf{\bibinfo{volume}{91}},
  \bibinfo{pages}{166102} (\bibinfo{year}{2003}).



\bibitem[{\citenamefont{Ishimoto and Kikuchi}(2005{\natexlab{a}})}]{IK051}
\bibinfo{author}{\bibfnamefont{Y.}~\bibnamefont{Ishimoto}} \bibnamefont{and}
  \bibinfo{author}{\bibfnamefont{N.}~\bibnamefont{Kikuchi}},
  \bibinfo{journal}{RIKEN-TH-49, cond-mat/0507477}
  (\bibinfo{year}{2005}{\natexlab{a}}).


\bibitem[{\citenamefont{Grosberg et~al.}(2002)\citenamefont{Grosberg, Nguyen,
  and Shklovskii}}]{GNS02}
\bibinfo{author}{\bibfnamefont{A.~Y.} \bibnamefont{Grosberg}},
  \bibinfo{author}{\bibfnamefont{T.~T.} \bibnamefont{Nguyen}},
  \bibnamefont{and} \bibinfo{author}{\bibfnamefont{B.~I.}
  \bibnamefont{Shklovskii}}, \bibinfo{journal}{Rev.\ Mod.\ Phys.}
  \textbf{\bibinfo{volume}{74}}, \bibinfo{pages}{329} (\bibinfo{year}{2002}).

\bibitem[{\citenamefont{Chaikin and Lubensky}(1995)}]{CL95}
\bibinfo{author}{\bibfnamefont{P.~M.} \bibnamefont{Chaikin}} \bibnamefont{and}
  \bibinfo{author}{\bibfnamefont{T.}~\bibnamefont{Lubensky}},
  \emph{\bibinfo{title}{Principles of condensed matter physics}}
  (\bibinfo{publisher}{Cambridge University Press},
  \bibinfo{address}{Cambridge}, \bibinfo{year}{1995}).

\bibitem[{T05(2005)}]{T05}
\bibinfo{journal}{in private communication with Y. Takenaka}
  (\bibinfo{year}{2005}).

\bibitem[{\citenamefont{Ishimoto and Kikuchi}(2006{\natexlab{b}})}]{IK05}
\bibinfo{author}{\bibfnamefont{Y.}~\bibnamefont{Ishimoto}} \bibnamefont{and}
  \bibinfo{author}{\bibfnamefont{N.}~\bibnamefont{Kikuchi}},
  \bibinfo{journal}{in preparation}  (\bibinfo{year}{2006}{\natexlab{b}}).

\end{thebibliography}
\end{document}